*Fast high-resolution metabolite mapping in the rat brain using $^1$H-FID-MRSI at 14.1T*

Dunja Simicic[1,2*], Brayan Alves[1,2*], Jessie Mosso[1,2,3], Guillaume Briand[1,2], Thanh Phong Lê[3], Ruud B. van Heeswijk[4], Jana Starčuková[5], Bernard Lanz[1,2], Antoine Klauser[6], Bernhard Strasser[7], Wolfgang Bogner[7], Cristina Cudalbu[1,2]

[1] CIBM Center for Biomedical Imaging, Switzerland

[2] Animal Imaging and Technology, École polytechnique fédérale de Lausanne (EPFL), Lausanne, Switzerland

[3] Laboratory of Functional and Metabolic Imaging, École polytechnique fédérale de Lausanne (EPFL), Lausanne, Switzerland

[4] Department of Diagnostic and Interventional Radiology, Lausanne University Hospital (CHUV) and University of Lausanne (UNIL), Lausanne, Switzerland

[5] Institute of Scientific Instruments of the CAS, Brno, Czech Republic

[6] Advanced Clinical Imaging Technology, Siemens Healthcare AG, Lausanne, Switzerland

[7] High-field MR Center, Department of Biomedical Imaging and Image-guided Therapy, Medical University Vienna, Vienna, Austria

*Joint first authors

**Abbreviations:**

MRSI - Magnetic Resonance Spectroscopic Imaging, FID - Free Induction Decay, AD - Acquisition Delay, TR - Repetition Time, SVD - Singular Value Decomposition, SNR - Signal to Noise Ratio, SD - Standard Deviation, FWHM - Full Width at Half Maximum, FT - Fourier Transform, NAA - N-Acetyl Aspartate, NAAG - N-acetylaspartylglutamate, Ins - myo-Inositol, Gln - Glutamine, Glu - Glutamate, GABA - Gamma-Aminobutyric Acid, Asp - Aspartic Acid, Ala - Alanine, Asc - Ascorbic Acid, Cr - Creatine, PCr - Phosphocreatine, tCr - total Creatine, Tau - Taurine, PCho - Phosphocholine, GPC - Glycerophosphocholine, PE - Phosphoethanolamine, Lac - Lactate, Glc - Glucose, GSH - Glutathione

Abstract:

Magnetic resonance spectroscopic imaging (MRSI) enables the simultaneous non-invasive acquisition of MR spectra from multiple spatial locations inside the brain. While $^1$H-MRSI is increasingly used in the human brain, it is not yet widely applied in the preclinical setting, mostly because of difficulties specifically related to very small nominal voxel size in the rat brain and low concentration of brain metabolites, resulting in low signal-to-noise ratio (SNR).

In this context, we implemented a free induction decay $^1$H-MRSI sequence ($^1$H-FID-MRSI) in the rat brain at 14.1T. We combined the advantages of $^1$H-FID-MRSI with the ultra-high magnetic field to achieve higher SNR, coverage and spatial resolution in the rat brain, and developed a custom dedicated processing pipeline with a graphical user interface for Bruker $^1$H-FID-MRSI: *MRS4Brain toolbox*.

LCModel fit, using the simulated metabolite basis-set and in-vivo measured MM, provided reliable fits for the data at acquisition delays of 1.30 ms. The resulting Cramér-Rao lower bounds were sufficiently low (<30%) for eight metabolites of interest (total creatine, N-Acetyl Aspartate, N-Acetyl Aspartate+N-acetylaspartylglutamate, total choline, glutamine, glutamate, myo-inositol, taurine), leading to highly reproducible metabolic maps. Similar spectral quality and metabolic maps were obtained with 1 and 2 averages, with slightly better contrast and brain coverage due to increased SNR in the latter case. Furthermore, the obtained metabolic maps were accurate enough to confirm the previously known brain regional distribution of some metabolites. The acquisitions proved high reproducibility over time.

We demonstrated that the increased SNR and spectral resolution at 14.1T can be translated into high spatial resolution in $^1$H-FID-MRSI of the rat brain in 13 minutes, using the sequence and processing pipeline described herein. High-resolution $^1$H-FID-MRSI at 14.1T provided robust, reproducible and high-quality metabolic mapping of brain metabolites with minimal technical limitations.

# 1. Introduction

Magnetic resonance spectroscopic imaging (MRSI) enables the simultaneous non-invasive acquisition of MR spectra from multiple spatial locations inside the brain. Even though such mapping of the metabolic regional differences in-vivo is very valuable for both clinical and preclinical research, the routine application of MRSI remains challenging due to several issues. These include long acquisition times, low signal-to-noise ratio (SNR), the need to develop in-house acquisition sequences and processing pipelines, the huge amount of data that needs to be handled, etc.[1] The availability of ultra-high magnetic fields (UHF) and the corresponding increase in SNR and spectral dispersion, combined with advanced pulse sequences and new encoding methods, improved the quality and speed of MRSI[2,3] specifically for human studies. At UHF, in the clinical setting, free induction decay ($^1$H-FID-MRSI) acquisitions are increasingly used[2,4]. FID-MRSI acquisition minimizes the $T_2$ relaxation and eliminates J-evolution. This in turn increases the SNR and potentially the number of detected metabolites. It also reduces chemical shift displacement errors and sensitivity to $B_0$ inhomogeneity[3,5–7]. Moreover, this simple sequence design enables a considerable acquisition time reduction by decreasing the repetition time (TR) while using an optimal Ernst's flip angle.

While $^1$H-MRSI is increasingly used in the human brain, it is not yet widely applied in the preclinical setting. However, preclinical disease models are very valuable since they allow an easy longitudinal follow-up of brain development, disease evolution and treatment efficacy. This is particularly evident in MR techniques where the field strengths of 7T and higher give access to unprecedented image and spectral resolution in the CNS deciphering an increased number of neurochemicals[8,9]. Therefore, establishing a fast and reliable MRSI protocol applicable to rat CNS could open new avenues for translational research. Specifically, it could allow important insight into diseases with complex regional distribution of metabolites where rat models are already widely studied. Rat and mouse MRSI techniques were previously used to study focal ischemia exploring metabolic differences and delineation of ischemic core and penumbra[10], and for characterization of glioblastoma and its response to treatment[11]. Furthermore, this technique has great potential in describing regional metabolic signatures of other diseases such as hepatic encephalopathy[12], multiple sclerosis[13,14], Alzheimer's disease[15], etc.

Limited applications of MRSI in preclinical studies come mostly because of difficulties specifically related to the small rat brain[8]. The resulting low SNR arises from the low concentration of brain metabolites combined with a very small nominal voxel size in rats (e.g. $0.75 \times 0.75 \times 2$ mm$^3$ in a $32 \times 32$ matrix)[10], while in the human brain the nominal voxel size remains fairly large even at high spatial resolution (e.g. $1.7 \times 1.7 \times 10$ mm$^3$, for a $128 \times 128$ matrix)[16]. Furthermore, there are additional challenges in terms of shimming of large volumes with many tissue interfaces, long measurement times with traditional MRSI sequences (e.g. 120 minutes)[10–12], water suppression artifacts and lipid contamination[8]. These are combined with the need to develop automatic and standardized processing pipelines, to perform quality assessment of a very large number of spectra and to estimate the precision and reliability of derived metabolite maps. To the best of our knowledge, at the present time, traditional phase-encoded MRSI using STEAM, PRESS or SPECIAL excitation schemes (at TEs between 2-10 ms), is still used for quantitative mapping of an extended number of metabolites[10,15,17] in preclinical settings, while no advanced processing and quality control pipelines are available for FID-MRSI Bruker data. Fast MRSI has been implemented for hyperpolarized molecular imaging, however under very different conditions and constraints (intense but quickly decaying signal, typically lower spatial resolution, very few metabolites with almost no overlap[18]).

In this context, we propose the implementation of $^1$H-FID-MRSI in the rat brain at 14.1T, something novel in the preclinical setting. We combined the advantages of $^1$H-FID-MRSI acquisitions with the UHF of 14.1T to achieve higher SNR and spatial resolution in the rat brain. Furthermore, as MRSI acquisitions are characterized by a very large amount of data that needs to be processed and subjected to quality control to reliably estimate and derive metabolite maps, we implemented a custom dedicated processing pipeline for Bruker $^1$H-FID-MRSI (*MRS4Brain toolbox*) able to perform water and lipid suppression, fitting, semi-automatic quality assessment, atlas-based segmentation, and an overlay of metabolic maps on the corresponding anatomical MRI image; all these are incorporated in a user-friendly toolbox with a graphical user interface (GUI).

## 2. Methods

All experiments were approved by The Committee on Animal Experimentation for the Canton de Vaud, Switzerland (VD 3022.1). Wistar male adult rats (n = 10 rats, 240 ± 50 g (estimated age of 8-9 weeks), Charles River Laboratories, L'Arbresle, France) under 1.5-2.5% isoflurane anesthesia were used. The body temperature of the animals was kept at 37.5 ± 1.0 °C by circulating warm water and measured with a rectal thermosensor. The respiration rate and body temperature were monitored using a small-animal monitor system (SA Instruments, New York, NY, USA). During the MRI scans, all animals were placed in an in-house-built holder, with their head fixed in a stereotaxic system using a bite bar and a pair of ear bars.

### *2.1 Two-compartment phantom and in-vivo $^1$H-MRSI acquisitions*

Fast $^1$H-FID-MRSI measurements were performed in the rat brain on a 14.1T horizontal magnet (Magnex Scientific, Yarnton, UK), a 1 T/m peak strength and 5500 T/m/s slew rate shielded gradient set (Resonance Research, Billerica, USA) interfaced to a Bruker console (BioSpec Avance NEO,

ParaVision 360 v1.1 and v3.3), and using a home-made transmit/receive quadrature surface coil (20 mm inner diameter).

$T_2$-weighted Turbo-RARE images were acquired in coronal and axial direction to position the MRSI slice for shimming, acquisition and for metabolite maps overlays (20 slices, TR = 3000 ms, $TE_{eff}$ = 27 ms, NA = 2, $RARE_{factor}$ = 6, 256 × 256 matrix, 0.8 mm slice thickness, FOV = 24 × 24 mm$^2$). A second $T_2$-weighted Turbo-RARE image was acquired in each rat for brain segmentation (60 slices, TR = 4100 ms, $TE_{eff}$ = 27 ms, NA = 10, $RARE_{factor}$ = 6, 128 × 128 matrix, 0.2 mm slice thickness, FOV = 24 × 24 mm$^2$).

For the high-resolution two-dimensional fast $^1$H-FID-MRSI acquisitions, we created a novel protocol in ParaVision 360 v1.1 and v3.3 based on the MRSI sequence ("CSI") provided by Bruker, knowing that no FID-MRSI protocol is provided. A slice-selective pulse-acquire sequence was used in combination with VAPOR[19] water suppression (slightly changed from Bruker implementation with hermite RF pulses, bandwidth: 600-660 Hz, flip angles 1 and 2: 84°/150°, last delay 22 or 26 ms, 614.7 ms duration) and 6 saturation slabs to minimize lipid contamination (slightly changed from Bruker "Fov Sat" implementation: 90° sech RF pulse, 1 ms, 20 kHz bandwidth, thickness of the slabs between 4-12 mm, 12.7 ms duration, Auto Spoiler (1.08 ms duration and 5.48% amplitude)) (Figure 1A and Supplementary Figure 1). The MRSI coronal slice was centered on the hippocampus, with 2 mm slice thickness and a FOV of 24 × 24 mm$^2$ (same FOV as for imaging, Figure 2). The matrix size was 31 × 31 leading to a nominal voxel size of 0.77 × 0.77 × 2 mm$^3$. The following acquisition parameters were used: spectral bandwidth of 7 kHz, 1024 FID data points, 140 µs dwell time, Cartesian $k$-space sampling, 8 dummy scans, acquisition delay (AD) 1.3 ms, phase encoding duration of 0.5 ms, navigator (gaussian 10° RF pulse, 1.37 ms duration, data size of 136, 23.86 ms module duration, stick to reference frequency) and drift compensation, chemical shift offset of -2 ppm (-2ppm from 4.7 ppm equal to 2.7 ppm working chemical shift), TR=813 ms leading to a total measurement time of 13 minutes. The excitation pulse was adjusted to the Ernst angle of 52° (0.5 ms calculated RF pulse with the Shinnar-Le Roux algorithm) using a mean metabolites $T_1$ relaxation time of ~1600 ms estimated at 14.1T[20]. First and second order shims were adjusted using Bruker MAPSHIM, first in an ellipsoid covering the full brain and further in a volume of interest (VOI) of 10 × 10 × 2 mm$^3$ centered on the MRSI slice. An in-plane Hamming $k$-space filter in the two spatial dimensions (built in PV 360 v1.1 and v3.3) was used to reduce the contribution of noisy high-frequency spatial components in image space (Supplementary Figure 2). A live demo on how to acquire in-vivo $^1$H-FID-MRSI datasets is posted on the webpage: [LIVE Demos – MRS4BRAIN - EPFL](#).

The following acquisitions were performed to validate the implementation of the fast $^1$H-FID-MRSI sequence and test the reproducibility of the acquired data:

- To test the precision of the ¹H-FID-MRSI sequence and its ability to separate metabolic profiles from different brain regions a two-compartment phantom was measured using ¹H-FID-MRSI and PRESS-MRSI sequences (for details see Supplementary Material).
- Seven in-vivo datasets with one and two averages, respectively, were acquired on n=6 rats.
- To test the reproducibility and stability of the acquired data, two additional in-vivo measurements were performed:
    - 3 acquisitions with 1 average and 2 acquisitions with 2 averages were performed in an interleaved mode in one rat. Before each scan, the water linewidth was measured and if needed a MAPSHIM was performed in the $10 \times 10 \times 2$ mm³ VOI. The achieved water linewidth was in the range of 24-30 Hz.
    - One rat was scanned twice at a 2 week-interval using 1 and 2 averages.
- The ¹H-FID-MRSI sequence was further improved by reducing the RF pulse and phase encoding durations to 0.2 and 0.3 ms, respectively, which led to an AD of 0.94 ms. Additional in-vivo acquisitions were performed with these new parameters using n = 4 additional rats (the results are provided in the Supplementary Material).

## 2.2 Processing pipeline, semi-automatic quality control and automatic brain segmentation

A homemade processing pipeline was developed in MATLAB specifically for ¹H-FID-MRSI datasets (Bruker, ParaVision 360 v1.X and v3.X formats). The pipeline was partially inspired by previous human brain ¹H-FID-MRSI processing pipelines[21] and contains the following steps:

1. Data formats: Bruker MRSI datasets already FT in the image space (*fid* for ParaVision 360v1.X and *fid_proc.64* for ParaVision 360v3.X) and further stored in 2D matrices, where each voxel contains one FID.
2. The water power mask was used to filter out the voxels located outside the brain and later for the lipid suppression (step 4 below). The power of the separately acquired water signal was computed by summing the squared magnitude of the frequency-domain signal. The segmentation condition for the water power mask was given by half the mean power over the full slice; a voxel was considered in the brain region if its power was higher than this threshold (Figure 1B).
3. Residual water signal removal from the metabolite signal using a Hankel-Singular Value Decomposition (HSVD) water suppression[21,22].
4. When no *k*-space Hamming filter is used, an SVD-based lipid suppression[21], based on the assumption that lipids and metabolites are orthogonal in the time or frequency and the spatial domain, can be applied. A brain/scalp segmentation was performed using the water power mask from step 2. The scalp voxels were subsequently used to generate an orthogonal basis for the lipid components by performing an SVD on the lipid voxel data. The rank of the basis was determined by the energy ratio between the brain and the scalp region after application of the projector: $E_{Brain}/E_{Skull} \geq \alpha$ (more information in[21]), with α being manually set by the user in the pipeline (α was set to 0.5 in the present study).

5. For an overall quality control of the acquired data the following maps were computed: a linewidth and $\Delta B_0$ map (i.e. frequency shifts) using the real spectrum of the water signal and an SNR map using the metabolite signal. The SNR was calculated from the resonance NAA at 2.01 ppm, defined as the NAA peak height in the real spectrum divided by one standard deviation (SD) of the noise measured in a noise-only region of the real part of the spectrum (from (-0.9) to (-1.25) ppm)[23] (Supplementary Figure 3). All these maps were then saved in the experiment folder.

6. LCModel (Version 6.3-1N) fitting with details provided below. Fitting was applied on voxels located inside the brain and selected using the water power mask from step 2.

7. Semi-automatic quality control (QC) after fitting: the pipeline uses the values of SNR (computed as described in step 5) and FWHM from LCModel. Fixed thresholds for quality filtering were chosen to be set above or equal to SNR=5[24,25] and below 125% of the average over the number of voxels $\overline{FWHM}$[8]. An additional acceptance criterion with respect to the Cramer-Rao lower bound of the metabolites located in the voxel was added for the in-vivo datasets: for each metabolite, the CRLB was defined to be lower or equal than 30% (Supplementary Figures 4 and 5).

8. Creation of metabolite maps overlaid to the corresponding MRI image based on the semi-automatic quality control performed after fitting. For visualization purposes, a bicubic convolution interpolation can be used on the metabolic maps.

9. An atlas based automatic segmentation tool was also implemented (Supplementary Figure 6): a multi-step registration was computed with an anatomical template, created for this purpose and based on the SIGMA atlas[26]. The transformation found with registration was applied to the labeled template and the regions can be reshaped to correspond to the MRSI spatial resolution. This segmentation tool also provides a full brain mask which can be used in the steps 2 and 4 above, replacing the water power mask.

The processing pipeline was incorporated in a MATLAB toolbox with a user-friendly GUI: *MRS4Brain toolbox* (Supplementary Figure 6) with the objective to provide a standardized processing tool for Bruker preclinical MRSI data.

*2.3 Fitting and quantification*

The spectra contained in the brain region (obtained by brain/scalp segmentation) were quantified using LCModel. The basis-set of metabolites was simulated using NMRScopeB (18 metabolites) from jMRUI[27], using published values of J-coupling constants and chemical shifts[28,29] and the pulse-acquire sequence with the same parameters as for the in-vivo $^1$H-FID-MRSI metabolite acquisitions (Figure

1C). Two basis-sets were created: one for AD=1.3 ms and the second for AD=0.94 ms. In each basis-set the following metabolites were included: alanine (Ala), aspartate (Asp), ascorbate (Asc), creatine (Cr), phosphocreatine (PCr), γ-aminobutyrate (GABA), glutamine (Gln), glutamate (Glu), glycerophosphocholine (GPC), glutathione (GSH), glucose (Glc), inositol (Ins), N-acetylaspartate (NAA), N-acetylaspartylglutamate (NAAG), phosphocholine (PCho), phosphoethanolamine (PE), lactate (Lac), taurine (Tau). PCho and GPC, and Cr and PCr were expressed only as tCho (PCho + GPC) and tCr (Cr+PCr) due to better accuracy in the estimation of their concentration as a sum.

For both ADs (1.3 ms and 0.94 ms), the corresponding macromolecule (MM) spectrum was acquired with the same sequence as for the metabolites. For that, a double inversion recovery module was programmed into the $^1$H-FID-MRSI sequence (TI = 2200/850 ms, 15 × 15 matrix size, 6 averages, TR = 3400 ms, *HS1_R20.inv* RF pulse of 2 ms duration, 2048 FID data points, 7 kHz acquisition bandwidth, RF excitation was performed with a 90° pulse due to increased TR). Due to lower SNR of the acquired MM, six (AD=0.94 ms) or seven (AD=1.3 ms) voxels were summed to obtain the final MM signal and the residual water was removed using AMARES[30]. The metabolite residuals were also removed with AMARES from jMRUI as previously described[23] (tCr (~3.88 ppm), Glu+Gln (~3.72 ppm), Ins (~3.50 ppm), Tau (~3.39 ppm), Tau+tCho (~3.21 ppm), NAA (~2.64 ppm), Glu (~2.32 ppm), Gln (~2.09 ppm)) and the final MM signal used in the basis-set is plotted in Figure 1C for AD=1.3 ms. A live demo on how the in-vivo acquired MM were processed is posted on the webpage: LIVE Demos – MRS4BRAIN - EPFL. tCr was used as a reference for our model due to the short TR used in the sequence and the $T_1$ relaxation of the metabolites. The control files used for LCModel quantification are provided with the *MRS4Brain toolbox*.

A detailed table of the acquisition and processing parameters following the experts' consensus recommendations on minimum reporting standards in in-vivo MRSI is presented in Supplementary Table 1.

*2.4 Data display and statistics*

All data are presented as mean ± SD. To evaluate the feasibility of the described $^1$H-FID-MRSI sequence to quantify previously known brain regional changes[26], an automatic segmentation of two different brain regions was performed (n = 4 rats, AD = 1.3 ms, 2 averages): a region with a mix of striatum and cortex and a region composed of hippocampus, as shown in Supplementary Figure 6. The average concentration in these brain regions was computed by averaging the concentration values over the number of voxels contained in the specific region after performing the semi-automatic quality control. Each regional mean concentration value is then averaged again over the number of measurements, giving estimates that are qualified as the mean of the mean values. Ins and Gln were chosen to be presented due to their known regional distribution in the rat brain[31]. Coefficient of variation maps (the ratio per voxel of one standard deviation over the mean of the relative concentration estimated after quantification) were computed for the three time points as well as for the two different averages and the two ADs.

Two-way analysis of variance (ANOVA) (implemented in the *MRS4Brain toolbox*) with respect to each metabolite in the neurochemical profile followed by Bonferroni's multi-comparisons post-test were performed. Two categorical factors were defined: the brain regions (hippocampus and cortex+striatum) and the number of averages (1 average and 2 averages). For the statistical tests with different AD, the analysis was conducted with the brain regions factor and the different AD (AD=1.3 ms and AD=0.94 ms). The significance level in two-way ANOVA was attributed as follows: *$p < .05$, **$p < .01$, ***$p < .001$, and ****$p < .0001$. All tests were two-tailed.

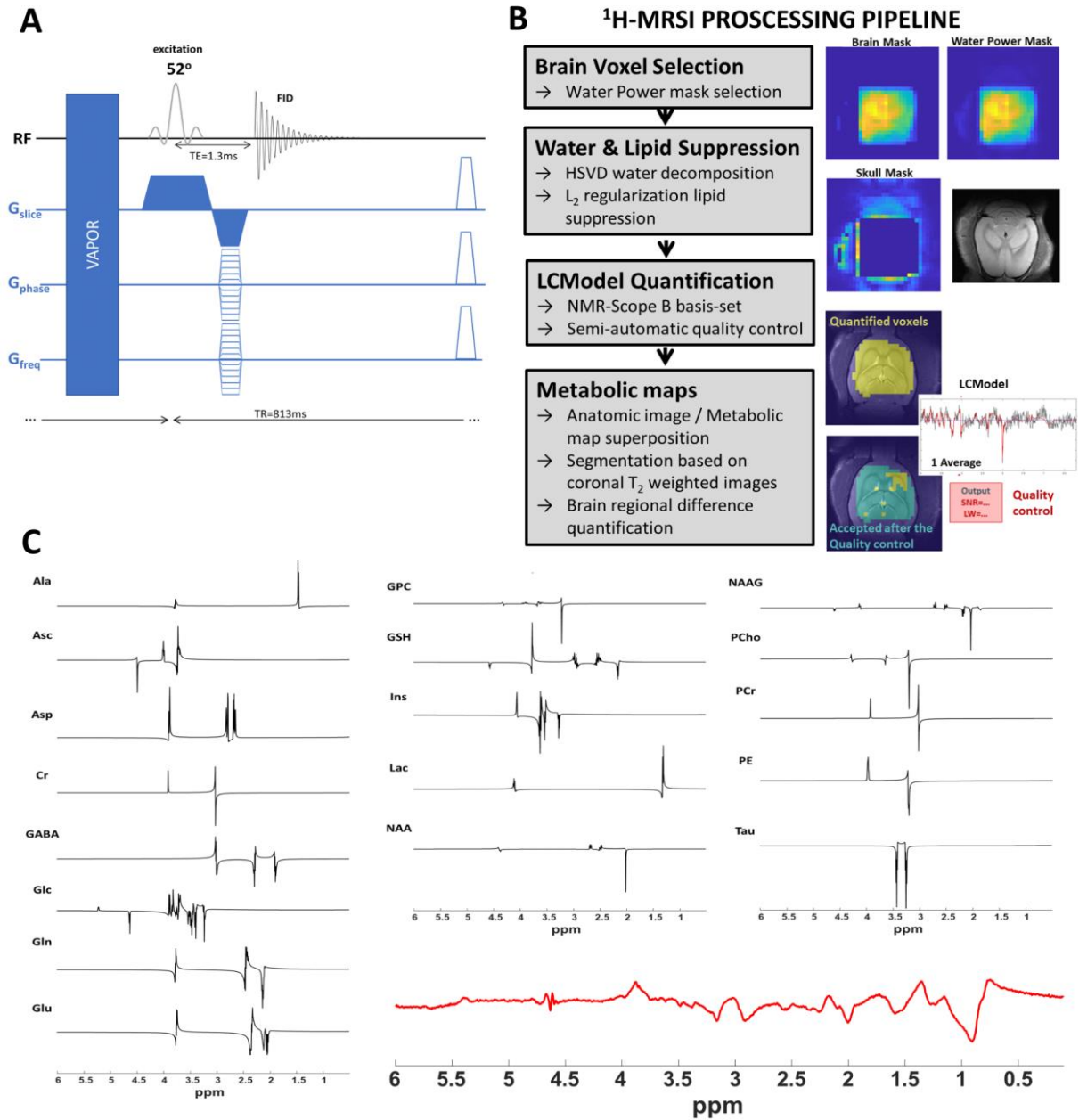

*Figure 1 A) A schematic drawing of the ¹H-FID-MRSI sequence used for data acquisition. B) Sketch of the processing pipeline used for ¹H-MRSI datasets C) The metabolites simulated using NMRScopeB (18 metabolites) from jMRUI and the in-vivo MM included in the basis-set.*

## 3. Results

The performance of the ¹H-FID-MRSI sequence was tested on a two-compartment phantom, in-vivo in the rat brain (n=10) using two different ADs with 1 and 2 averages (the results obtained with AD=0.94 ms are presented in the Supplementary Material Figure 7 and Table 4). In addition, some reproducibility tests were also performed in two rats with a total of 9 measurements with 1 and 2 averages.

## 3.1 Data quality

The shim adjustments using MAPSHIM, first in an ellipsoid covering the full brain then in a voxel of $10 \times 10 \times 2$ mm$^3$ centered on the MRSI slice, proved to be efficient when shimming in large areas. The average water linewidth over the 10 rats was $28.5 \pm 2.5$ Hz (measured in the $10 \times 10 \times 2$ mm$^3$ VOI centered on the MRSI slice) and the global linewidth estimated by LCModel, calculated from the lineshape convolution[32] ranged from 10 to 32 Hz (estimated after the QC step no 7; good quality spectra linewidths are shown in Figure 2), respectively. This translated into good quality spectra (e.g. high spectral resolution obtained with efficient shimming, efficient water and lipid suppression, flat baseline) in a large number of nominal voxels in the MRSI matrix after the application of the water power mask (step 2 in the MRSI processing pipeline, ~230 voxels selected by the water power mask (~25% of the whole slice)). Therefore, a large in-plane coverage was achieved extending also towards the edges of the brain and was not limited to a rectangular volume like for PRESS-MRSI. The computed water maps from the two-compartment phantom confirmed the increased in-plane coverage when using the $^1$H-FID-MRSI sequence (Supplementary Figure 8). Moreover, the water maps showed that the acquired signal comes from two clearly separated compartments, while the visual comparison of metabolite spectra from different positions in the MRSI matrix highlighted the ability of the $^1$H-FID-MRSI sequence to separate metabolic profiles from two closely positioned compartments with different metabolites.

The overall quality assessment computed by step 5 in the MRSI processing pipeline is shown in Supplementary Figure 3, highlighting the quality of the acquired data after automatically filtering out the spectra located outside the brain based on the water power mask (step 2 in the MRSI processing pipeline). As can be seen, the frequency shifts over the investigated MRSI slice (i.e. $\Delta B_0$) were overall below 20 Hz, while they were increased in some parts towards the edges of the MRSI slice. These small water frequency shifts highlighted a good quality shimming as shown in the same figure by the water linewidth map (around 20-30 Hz with broader linewidths in regions with higher $\Delta B_0$). Furthermore, small water frequency shifts are also mandatory for good water suppression during the acquisition of the MRSI datasets. The region containing broader linewidths was further removed by the semi-automatic quality control (step 7 in the MRSI processing pipeline) as shown by the smaller voxel selection afterwards. Despite the small nominal voxel size ($0.77 \times 0.77 \times 2$ mm$^3$) we reproducibly obtained good SNR data. The computed NAA SNR map after the step 5 in the MRSI processing pipeline (Supplementary Figure 3A) resulted in an averaged SNR value, over the entire datasets contained in the power mask (without QC, thus SNR<5 also selected), of $13.1 \pm 2.2$ for all the measurements reported as mean of the mean values of each rat (n=10 rats) performed with 1 average (and an average SNR value of $16.1 \pm 2.9$ with 2 averages, n=7 rats), while the individual spectra selected by the QC (step 7 in the MRSI processing pipeline) displayed SNRs ranging between 5 to 35 depending on the acquisition

(lower SNRs always towards the margins of the metabolite maps due to surface coil $B_1$ coverage and potentially broader linewidths).

Figure 2 shows an example of spectra acquired with 1 and 2 averages from two distinct locations in the rat brain (left and right hippocampus) and the corresponding LCModel fits with the corresponding SNR (estimated during QC step 5) and LCModel estimated linewidths. High-quality spectra were acquired with similar spectral quality between 1 and 2 averages and between right and left hippocampus, without baseline distortions due to water residuals or lipid contamination. The averaged metabolite SNR values computed over the entire datasets contained in the power mask increased by ~30% for the measurements performed with 2 averages. This SNR increase is an estimate as the data after the power mask selection still needs to pass the quality control (step 7 in the MRSI processing pipeline). Thus, low quality spectra can still be present for this specific estimation of averaged metabolite SNR as highlighted in Supplementary Figure 3B.

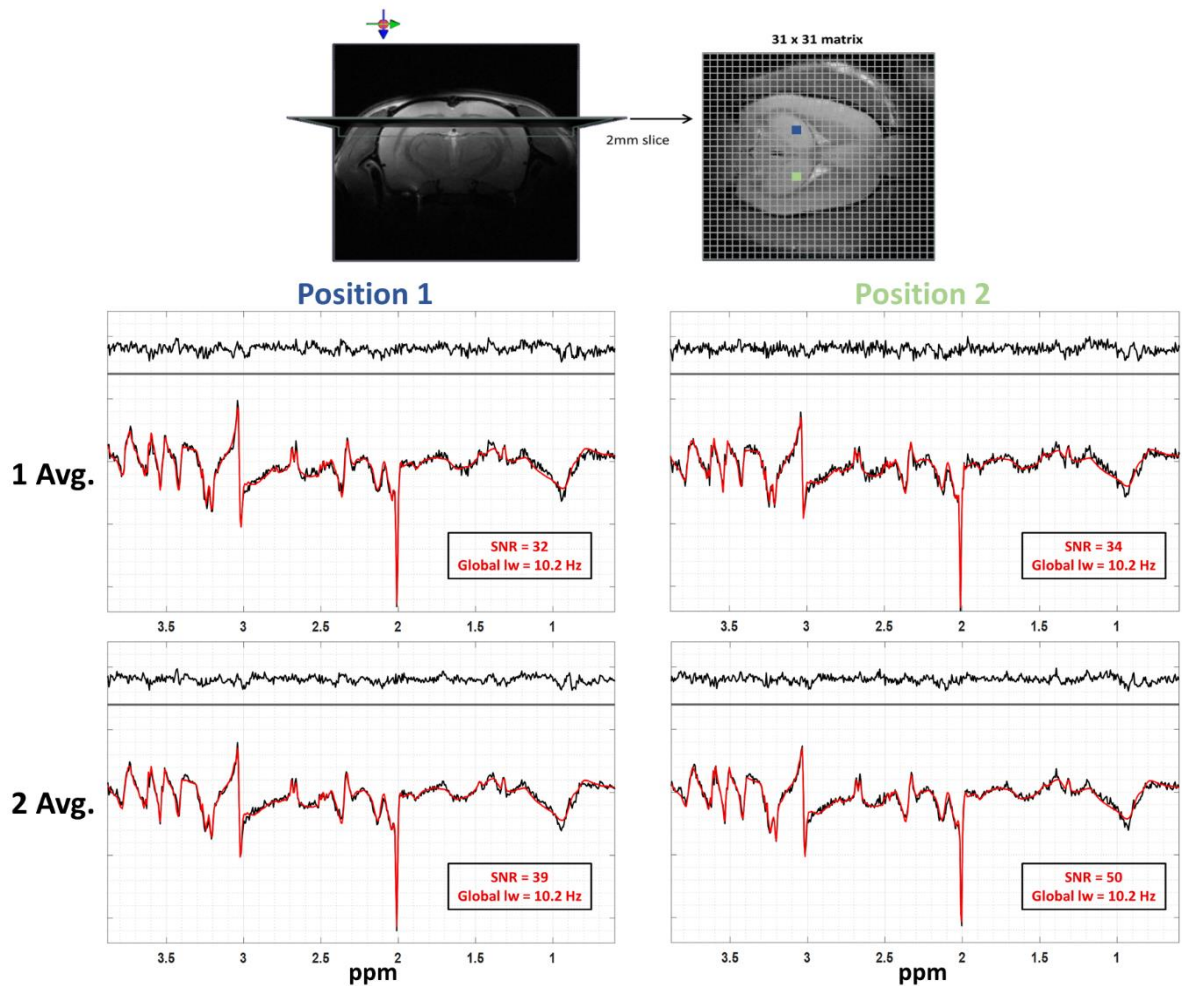

*Figure 2 Examples of the MRSI slice position in both axial and coronal views, and the spatial resolution are shown on the upper part of the figure. The high-quality of spectra and of the LCModel fits are shown for spectra from two different positions in the MRSI matrix with one and two averages (AD=1.3 ms). SNR was calculated using NAA peak height in the real spectrum divided by one standard deviation (SD) of the noise measured in a*

*noise-only region of the real part of the spectrum (from (-0.9) to (-1.25) ppm), while the linewidth (lw) was estimated by LCModel.*

3.2 In-vivo metabolic maps

The LCModel fit, using the simulated metabolite basis-set and in-vivo measured MM, provided reliable fits for the data obtained with both one and two averages at AD=1.3 ms (Figure 2, Supplementary Figure 9). The resulting Cramér-Rao lower bounds (CRLB's) were sufficiently low (<30%) for eight metabolites of interest (tCr, NAA, tNAA, tCho, Gln, Glu, Ins, Tau) leading to reproducible metabolic maps (examples of CRLB maps for NAA, tCho, Ins, are available in Supplementary Figures 4 and 5). The mean value of CRLBs over the brain regions (whole slice) are reported in Supplementary Table 2 and 3. The metabolic maps overlaid on the corresponding anatomical image for NAA, Ins and tCho are shown in Figure 3 for 1 and 2 averages after the application of the semi-automatic quality control described in the MRSI processing pipeline (step 7). Although the metabolic maps obtained from the acquisition with two averages provided a slightly better contrast and brain coverage due to increased SNR, the maps kept the same pattern when using one average proving that this very fast acquisition leads to a satisfactory output. As can be seen in Supplementary Figure 3B, the semi-automatic quality control led to the removal of some voxels in some areas where the quality of the spectra did not pass the selection criteria, leading to a slightly different coverage than the one first predicted by the water power mask (64% and 73% of the voxels in the water power mask were accepted for 1 and 2 averages, respectively). CRLB maps for these three metabolites can be seen in Supplementary Figure 4. Table 1 illustrates the mean concentration estimates on 6 rats (7 measurements) in two different brain regions. The results obtained with 1 and 2 averages are highly reproducible in the investigated brain regions with 1-6% statistically non-significant differences between 1 and 2 averages for the reported metabolites. tCr was used as internal reference due to the $T_1$ weighting induced by the short repetition time used in the current study.

To illustrate the ability of the described $^1$H-FID-MRSI sequence and processing pipeline to quantify previously known brain regional changes, Figure 4 depicts the obtained metabolic maps of Ins and Gln, as examples, which were accurate enough to confirm the previously known[31] brain regional distribution of these metabolites. Of note, due to the MRSI slice position we report two brain regions pooled together, thus the brain regional differences for some metabolites showing opposite differences in these two brain regions can be decreased or canceled out. We quantified up to 34 region-specific nominal voxels in the hippocampus and 30 for cortex+striatum. Furthermore, Figure 4C confirms the reproducibility of our MRSI acquisition displaying the Ins and Gln concentration ratios to tCr obtained from seven different rats, proving the reproducible estimation of these metabolites even with one average (13 minutes).

The *MRS4Brain toolbox* with a user-friendly MATLAB-based interface combining all processing steps was developed and used in the current study (Supplementary Figure 6), taking into account the huge amount of data acquired during each scan and the specific needs regarding the processing, artifacts removal when possible, fitting, quality control and display of reliably segmented metabolic maps. Each step of this processing pipeline was described and the results presented herein. The duration of the processing and quantification was estimated to be 10 minutes per dataset.

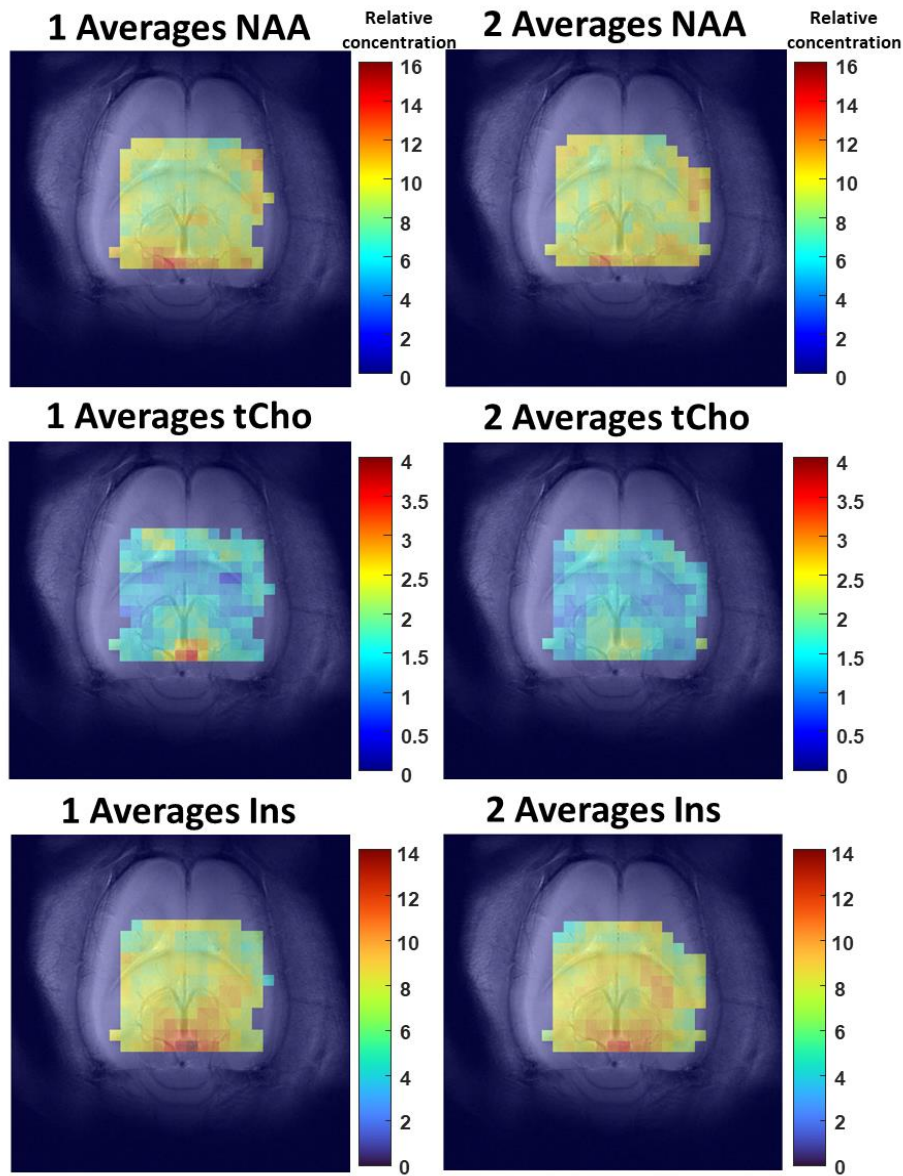

*Figure 3*. Representative metabolic maps obtained from the LCModel quantification results of the data acquired in one rat with one average (on the left) and two averages (on the right) with QC applied (AD=1.3 ms). As can be seen, the obtained metabolic maps are similar between 1 and 2 averages, with a slightly better contrast for 2 averages due to an increased SNR. The brain coverage is dictated by the position of the saturation slabs, by the position and coverage of the quadrature surface coil and by the QC criteria. The metabolic maps were superimposed to the corresponding anatomical image using the MRS4Brain toolbox. The scales correspond to LCModel outputs when referenced to tCr by setting its concentration to 8 mmol/kg$_{ww}$. No interpolation was used on the metabolite maps.

| */tCr [mmol/kg_ww] | 1 Averages | | */tCr [mmol/kg_ww] | 2 Averages | |
|---|---|---|---|---|---|
| | Hippocampus | Striatum + Cortex | | Hippocampus | Striatum + Cortex |
| NAA | 9.46 ± 0.38 | 9.50 ± 0.42 | NAA | 9.62 ± 0.32 | 9.81 ± 0.39 |
| Gln (**) | 5.15 ± 0.41 | 5.91 ± 0.26 | Gln (**) | 5.25 ± 0.38 | 6.27 ± 0.26 |
| Glu (**) | 11.00 ± 0.39 | 11.93 ± 0.44 | Glu (**) | 11.18 ± 0.46 | 11.74 ± 0.38 |
| Ins (**) | 9.50 ± 0.35 | 7.46 ± 0.35 | Ins (**) | 9.48 ± 0.32 | 7.72 ± 0.36 |
| Tau | 7.09 ± 0.26 | 6.92 ± 0.31 | Tau | 7.08 ± 0.23 | 7.04 ± 0.28 |
| GPC+PCho | 1.76 ± 0.09 | 1.85 ± 0.11 | GPC+PCho | 1.68 ± 0.10 | 1.86 ± 0.11 |
| NAA+NAAG | 10.86 ± 0.49 | 10.94 ± 0.50 | NAA+NAAG | 10.79 ± 0.41 | 11.16 ± 0.43 |

*Table 1: Quantitative reproducibility assessment in two brain regions using 1 and 2 averages (mean of the mean over 7 measurements in 6 rats, per average) at AD=1.3 ms. No significant differences between 1 and 2 averages were found, however brains regional differences are marked with \*,\*\*.*

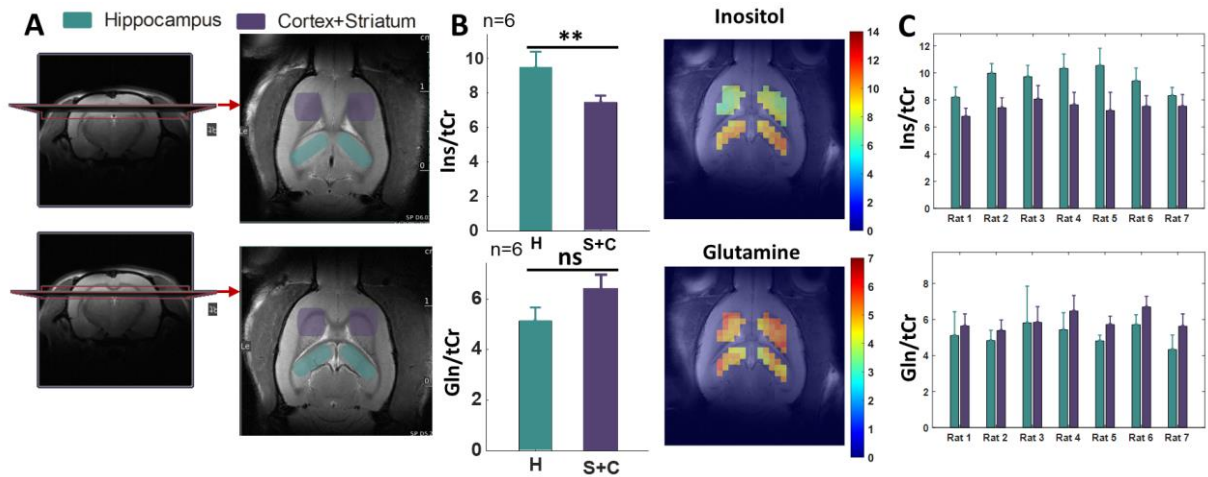

*Figure 4: A) Position of the MRSI slice in both axial and coronal views with manually marked brain regions within the slice for illustration purposes only. B) Brain regional differences in resulting concentrations of Ins and Gln (n=4 rats using the quantifications of the MRSI data acquired with 2 averages, AD=1.3 ms, on the left) derived from the metabolic maps highlighting the automatic segmentation of the two brain regions (in-vivo example, on the right). C) Reproducibility of metabolite quantifications (Ins and Gln) between the 6 different animals (7 acquisitions) in the MRSI acquisitions with 1 average (AD=1.3 ms). The semi-automatic quality control was applied. No interpolation was used on the metabolite maps.*

3.3 In-vivo reproducibility

High reproducibility was obtained for the measurement of brain metabolites using [1]H-FID-MRSI, as proposed in our study. The good quality shimming was also kept during the reproducibility studies, with water linewidth ranging between 24-26 Hz over the entire duration of the scans (total of 4 hours, Figure 5), while for the rat scanned at 2 weeks interval the water linewidth ranged between 25-28 Hz for both measurements. Figure 5 illustrates qualitatively the high reproducibility of metabolic mapping in three repeated measurements with 1 average for NAA, tCho and Ins over 4 hours. On average, the voxel-to-voxel coefficients of variation (CV) over the entire slice (Figure 5, Supplementary

Figure 10) were 6 ± 4% for both NAA and Ins, with 4 ± 4% and 3 ± 4% for NAA and with 5 ± 5% and 4 ± 5% for Ins on the hippocampus and striatum+cortex regions respectively. For tCho, average CVs over the entire slice were 15 ± 9%, with 13 ± 11% on the hippocampus region and of 6 ± 8% on the striatum+cortex region. Coefficient of variation maps for the three time points as well as for the two different averages and the two ADs can be found in Supplementary Figure 10.

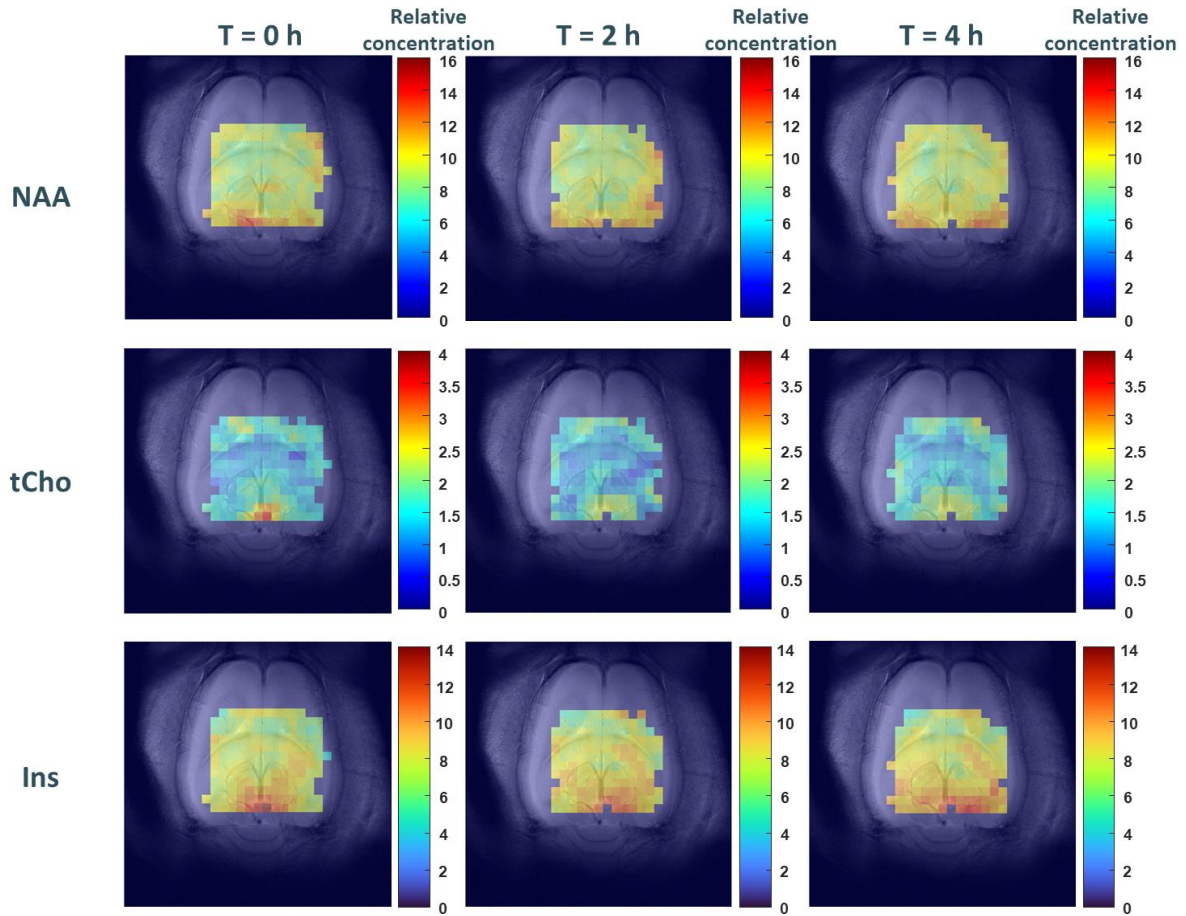

*Figure 5: Qualitative reproducibility of metabolic maps illustrated for one experiment where three $^1$H-FID-MRSI measurements were performed with 1 average at AD=1.3 ms in the rat brain during 4 hours of acquisition with QC applied. As can be seen the obtained metabolic maps are reproducible over the 4 hours of acquisition. The brain coverage is dictated by the position of the saturation slabs, by the position and coverage of the quadrature surface coil and by the QC criteria. The scales correspond to LCModel outputs when referenced to tCr by setting its concentration to 8 mmol/kg$_{ww}$. No interpolation was used on the metabolic maps. The metabolite maps were superimposed to the corresponding anatomical image using the MRS4Brain toolbox.*

## 4. Discussion

We presented the first implementation and validation of fast $^1$H-FID-MRSI in the rat brain at 14.1T with an increased brain coverage, accurate quantification results and robust metabolite maps which enabled us to highlight the brain regional distribution of some metabolites in 13 minutes. We showed that our results are reproducible, providing a sound basis for a wider application of $^1$H-FID-

MRSI in a preclinical setting. This was made possible by the use of several technical improvements: implementation of the sequence with a total acquisition time of 13 minutes (i.e. short TR and AD), high quality shimming, water and lipids suppression combined with an in-house developed MRSI processing pipeline specifically designed for such studies. *MRS4Brain toolbox* is a new MRSI processing pipeline developed for Bruker MRSI data containing several processing steps combined with semi-automatic quality control and segmentation tool. Our pipeline eliminated the need to manually classify the data by expert users, to manually overlay the metabolic maps on the anatomical image and to manually perform brain segmentation concomitantly on anatomical MR images and metabolic maps, which is time-consuming (a minimum of 961 spectra needs to be assessed for each scan), and subjective potentially lowering reproducibility.

Ultra-short AD values are particularly useful for the quantification of J-coupled metabolites. In our study using an AD=1.3 ms no J-modulation was observed, which simplified spectral quantification substantially, and increased the SNR of J-coupled metabolites[5]. Another advantage of FID-MRSI sequences is the decreased chemical shift displacement artifact, which is increasing proportionally to $B_0$, being particularly large for narrow-band, slice-selective RF pulses. As no refocusing pulses are required for FID-MRSI sequences, there is no in-plane chemical shift displacement artifact. At 7T in the human brain, it has been shown that the use of a short, slice-selective excitation RF pulse with high bandwidth and reduced flip-angle reduced the chemical shift displacement artifact to only ~5% per part per million[5].

MAPSHIM combined with our shimming protocol proved to be efficient leading to an increased brain coverage (not limited to a standard rectangular volume). This resulted in good quality and reliable metabolite maps for a number of metabolites (NAA, tNAA, Glu, tCho, Ins, Gln, Tau, etc.). However, as expected a better quality of the acquired MRSI datasets was observed in the center of the MRSI slice, as the usage of surface coils leads to a quality reduction at the edges of the MRSI slice due to the coil sensitivity profile. Furthermore, it is worth pointing towards the necessity of having automatic quality control steps implemented in a processing pipeline for MRSI datasets, as those presented herein, first for a fast and automatic elimination of low-quality spectra (i.e. spectra located outside the brain) and also for an overall quality control of the acquired data, as "bad" quality data can be excluded directly without spending time in further processing them (Supplementary Figure 3).

The quantification of low-concentration metabolites (i.e. GABA, Asp, PE, Asc, GSH) was not shown due to the smaller number of voxels reliably quantified and thus needing further improvements, either by further reducing the AD and/or by increasing the number of averages. Ala and Lac were not reported due to possible lipid contamination/lipid removal even though overall good lipid suppression was obtained for the [1]H-FID-MRSI datasets. In human studies at 7T, different groups have addressed this issue for [1]H-MRSI, using OVS or saturation slabs, inversion recovery, frequency-selective suppression or suppression based on $B_1$ shimming approaches or by improving the point-spread function by increasing the matrix size and applying spatial Hamming filtering[5,6,33–36]. In our study, lipid suppression was performed by saturation slabs during the acquisition and enhanced either by applying a spatial Hamming filter or by the usage of the SVD-based lipid suppression[21]. As saturation slabs were also used during the acquisition of the MRSI datasets, the α parameter in the SVD-based lipid suppression was set manually to 0.5. The use of a spatial Hamming filter was necessary mainly to reduce the impact of noisy high-frequency spatial components of the signal, which are further leading to higher spectral noise and thus to spatial fluctuations in the metabolite maps when all $k$-space spatial components are equally weighted (i.e. Hamming filter is not used, see Supplementary Figure 2). Importantly, the Hamming filter also reduces the lipid contamination by reducing spatial signal spread with a less extended point-spread function[25,37], with a drawback of increased nominal voxel size from 1.21 to 1.86 (Supplementary Figure 11). Therefore, the combination between $k$-space filtering and the SVD-based lipid suppression should be tested with care as it could lead to distortions in the MM pattern. As such, the two techniques were not combined in the present study.

The short TR used in our study led to short acquisition times (i.e. 13 minutes for a 31 × 31 MRSI matrix) but also to stronger $T_1$ weighting. Adjusting the Ernst excitation flip angle led to a reduction in the $T_1$ weighting as previously shown at 7T in the human brain[5]. Furthermore, we used tCr as internal reference during the quantification step as metabolite ratios are less sensitive to flip angle errors and $T_1$ weighting. Of note, when correcting for $T_1$ relaxation times differences in metabolite $T_1$ relaxation times and the different $T_1$ values of resonances originating from the same molecule (i.e. NAA and $CH_2$ groups[20]) should be taken into account.

The current manuscript demonstrates the feasibility of $^1$H-FID-MRSI in the rat brain. The sequence, acquisition protocol and processing pipeline can be translated to the mouse brain without any difficulty, as traditional phase encoded PRESS and SPECIAL acquisition have been performed in the mouse brain[11,38]. If needed the voxel size can be adapted by decreasing the FOV and averaging can be performed for the metabolite datasets.

**Limitations & Future steps**

The brain coverage was limited by the coverage of the 2-loop surface RF coil used with potential $B_1$ inhomogeneities towards the edges of the metabolic maps, and by variable $B_0$ homogeneity in some brain regions leading to metabolite maps which do not fully cover the brain as in human studies. As such, future studies will focus on improving both the RF excitation homogeneity and RF reception coverage, potentially by using surface transmit/receive phased arrays, or by combining surface-receive with volume-transmit RF coils, whose compatibility regarding the power deposition in $^1$H-FID-MRSI at high duty cycle needs to be evaluated. The increased SNR and brain coverage would be potentially beneficial also for static field shimming. The use of saturation slabs might also contribute to a decreased brain coverage if placed too close to the brain and with suboptimal transition bands. Supplementary Figure 1 illustrates a new version of the saturation slabs with potential for increased brain coverage. The approach is based on thinner saturation slabs with corresponding sharper transition bands, as well as the use of pairs of thinner overlapping bands in outer areas where extended signal cancellation is required. This leads to better transition bands for the saturation slabs and thus less signal suppression in the brain tissue. By further reducing the AD to ~0.7 ms we would also be able to slightly increase the SNR and thus the brain coverage. A further approach is to predict the missing FID points due to the acquisition delay as shown previously[7,21] and consequently eliminate the 1$^{st}$ order phase distortions. In parallel, further reducing the TR and consequently adapting the water and lipid suppression efficiency for shorter TRs together with the implementation of a concentric rings encoding[1,39] would allow faster acquisitions and potentially enable 3D encoding[40]. Future studies will also investigate the efficiency of lipid suppression using metabolite-lipid orthogonality and the best compromise between the amount of lipid contamination (depending on the RF coils used and presence or not of saturation slabs in the sequence) and the rank of the basis for the approximated lipid subspace. In the present study, tCr was used as an internal reference. In the next step we will focus on correcting for $T_1$ relaxation times using previously published values[20]. Some metabolites showed higher concentration towards the edges of the maps, which could be due to some technical issues (surface coils induced $B_1$ inhomogeneities, lower SNR, fitting issues, etc.).

## 5. Conclusion

In this study, we demonstrated that the increased SNR and spectral resolution at 14.1T can help achieve high spatial resolution in $^1$H-FID-MRSI of the rat brain in 13 minutes, using the sequence and processing pipeline (*MRS4Brain toolbox*) described herein. High reproducibility was achieved in several brain regions together with the detection of brain regional differences, as illustrated by the in-vivo measurements. As such, high-resolution $^1$H-FID-MRSI at 14.1T provided robust, reproducible and high-quality mapping of brain metabolites with significantly reduced technical limitations The short duration of the metabolite acquisition (i.e. 13 minutes) can be used for biomedical applications in different animal models where brain regional evolution of metabolites is questioned during disease evolution. Doing so, preclinical $^1$H-FID-MRSI could provide insights into the pathogenesis of major neurological diseases and improve the understanding of the basic neurochemical mechanisms involved in brain metabolism.

## Data availability:

The *MRS4Brain toolbox* version v.0.1 is available on the following repository: https://github.com/AlvBrayan/MRS4Brain-toolbox. Experimental data used in the present manuscript are also available on the same repository. Finally, a live demo on how to acquire in-vivo $^1$H-FID-MRSI datasets and how to process in-vivo acquired MM is posted on the webpage: LIVE Demos – MRS4BRAIN - EPFL.

## Acknowledgments:

This work was supported by the Swiss National Science Foundation award n° 201218 and 207935 and by the Center for Biomedical Imaging of the UNIL, UNIGE, HUG, CHUV, EPFL, the Leenaards and Jeantet Foundations. We acknowledge the core facility ISI-MR, co-funded by the Czech-BioImaging large RI project (LM2023050 funded by MEYS CR), for the technical support with jMRUI. The authors thank Dr. Katarzyna Pierzchala for building the phantom and the veterinary staff at CIBM MRI EPFL AIT for support during experiments.

## Declaration of Conflict of Interest:

The authors have no conflict of interest to declare.

# Supplementary Material

**Methods: two-compartment phantom**

To test the precision of the $^1$H-FID-MRSI sequence and its ability to separate metabolic profiles from different brain regions we built a two-compartment phantom. The phantom consisted of two syringes; a small syringe (5 ml) positioned inside a big syringe (50 ml). The small syringe was filled with a high concentration (~50 mM) of Cr with 1 μmol/ml gadolinium (Gd, Diethylenetriaminepentaacetic acid gadolinium(III) dihydrogen salt hydrate, Sigma-Aldrich, 381667) dissolved in phosphate buffer saline (PBS) and the big syringe was filled with a high concentration (~40 mM) of Glu with 1 μmol/ml Gd dissolved in PBS. Both $^1$H-FID and PRESS-MRSI acquisitions were performed with the following parameters:

- PRESS-MRSI: TE = 12.7 ms, TR = 2000 ms, one average, 31 × 31 matrix in a FOV = 24 × 24 mm$^2$ leading to a total scan time of 32 minutes. The VOI of 10 × 10 × 2 mm$^3$ was centered in the phantom (positioned in the same slice used for $^1$H-FID-MRSI).
- $^1$H-FID-MRSI: acquisition delay AD = 1.3 ms, TR = 812 ms, one average, matrix size 31 × 31 in a FOV = 24 × 24 mm$^2$ (slice positioned to contain the PRESS VOI) leading to a total acquisition time of 13 minutes.

**Results: in-vivo AD = 0.94 ms**

To further minimize first-order phase problems and signal loss due to AD, we have implemented a new protocol where the AD was reduced at 0.94 ms by decreasing the duration of the excitation RF pulse (from 0.5 ms to 0.2 ms) and phase encoding (from 0.5 ms to 0.3 ms). An example of data acquired with this shorter AD is shown in Supplementary Figure 7 (corresponding CRLB maps in Supplementary Figure 5). Qualitatively, reproducibility of metabolite mapping was obtained between AD = 1.3 ms and 0.94 ms, as can be seen when comparing the metabolic maps of NAA, tCho and Ins. An approximate 9% increase in SNR was observed when comparing the averaged metabolite SNR maps between the two ADs (n = 4 rats for each AD). The estimated concentrations for 7 metabolites are displayed in Supplementary Table 4 for the two ADs and two brain regions. On average 1-16 % difference was measured between the two ADs, with a trend of decreased concentrations for AD of 0.94 ms. These differences were statistically significant for Ins (13%** hippocampus and 16%* cortex+striatum), tCho (16%** hippocampus and 16%* cortex+striatum) and Tau (9%*, only for the hippocampus). Several reasons can be at the root of these differences (different 1$^{st}$ order phase evolution due to different ADs which might impact LCModel fitting, even though the 1$^{st}$ order phase was simulated in the metabolites basis-set and fixed to zero in the LCModel control file; MM fitting, etc). Further investigations are needed to probe this aspect, not yet explored in the literature for FID-MRSI. This reduction in AD is potentially beneficial for low concentration and J-coupled metabolites as the 1$^{st}$ order phase evolution will be smaller and less $T_2^*$ evolution will occur, which might also explain the differences in metabolite estimates at different ADs.

**Supplementary Figure 1:** Brief illustration and description of the saturation slabs: 1) used in the current study (Option 1); 2) Option 2 presents a new version of the saturation slabs with potential for increased brain coverage. The approach is based on thinner saturation slabs with corresponding sharper transition bands, as well as pairs of thinner overlapping slabs where extended signal cancellation is required, such as the slices 2&3. This leads to better transition bands and thus less signal suppression in the brain tissue.

1): Saturation slabs – Option 1 – used in the current manuscript

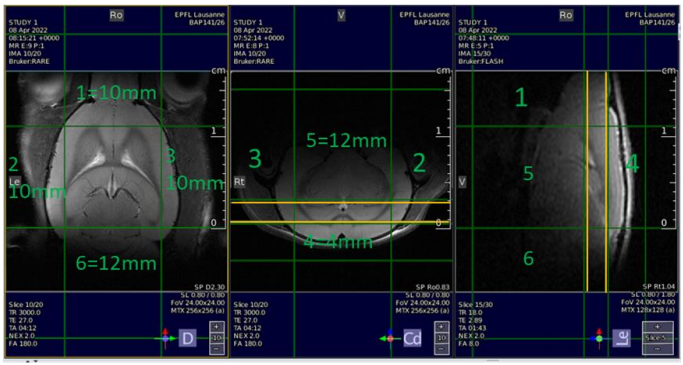

6 Saturation slabs
Thk between 4-12mm
2 mm slice thk for FID-MRSI

2) Saturation slabs – Option 2 – for an increased brain coverage

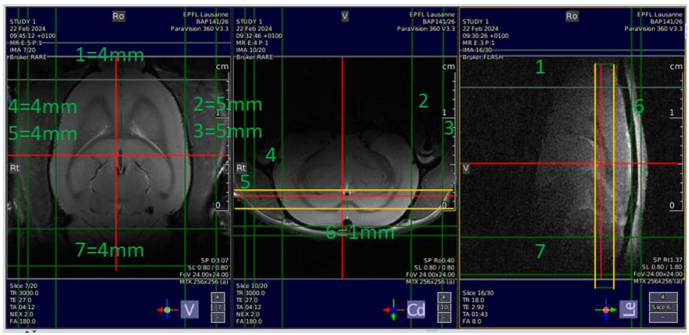

7 Saturation slabs
Thk between 1.5-4mm
Slabs 2&3 and 4&5 saturate the same region
2 mm slice thk for FID-MRSI

**Supplementary Figure 2:** Representative metabolic map of Ins obtained from the LCModel quantification results of the data acquired in one rat (1 average) with (on the left) and without (on the right) Hamming filter applied, with QC applied. Differences in coverage and metabolite distribution patterns can be observed between the two maps. Please note the presence of strong spatial fluctuations in metabolite concentrations in the non-filtered metabolite map, reflecting a significant inclusion of noise-dominated high-frequency spatial components in the image space. A representative spectrum from each set is plotted (black) with the respective fit (red) and residuals.

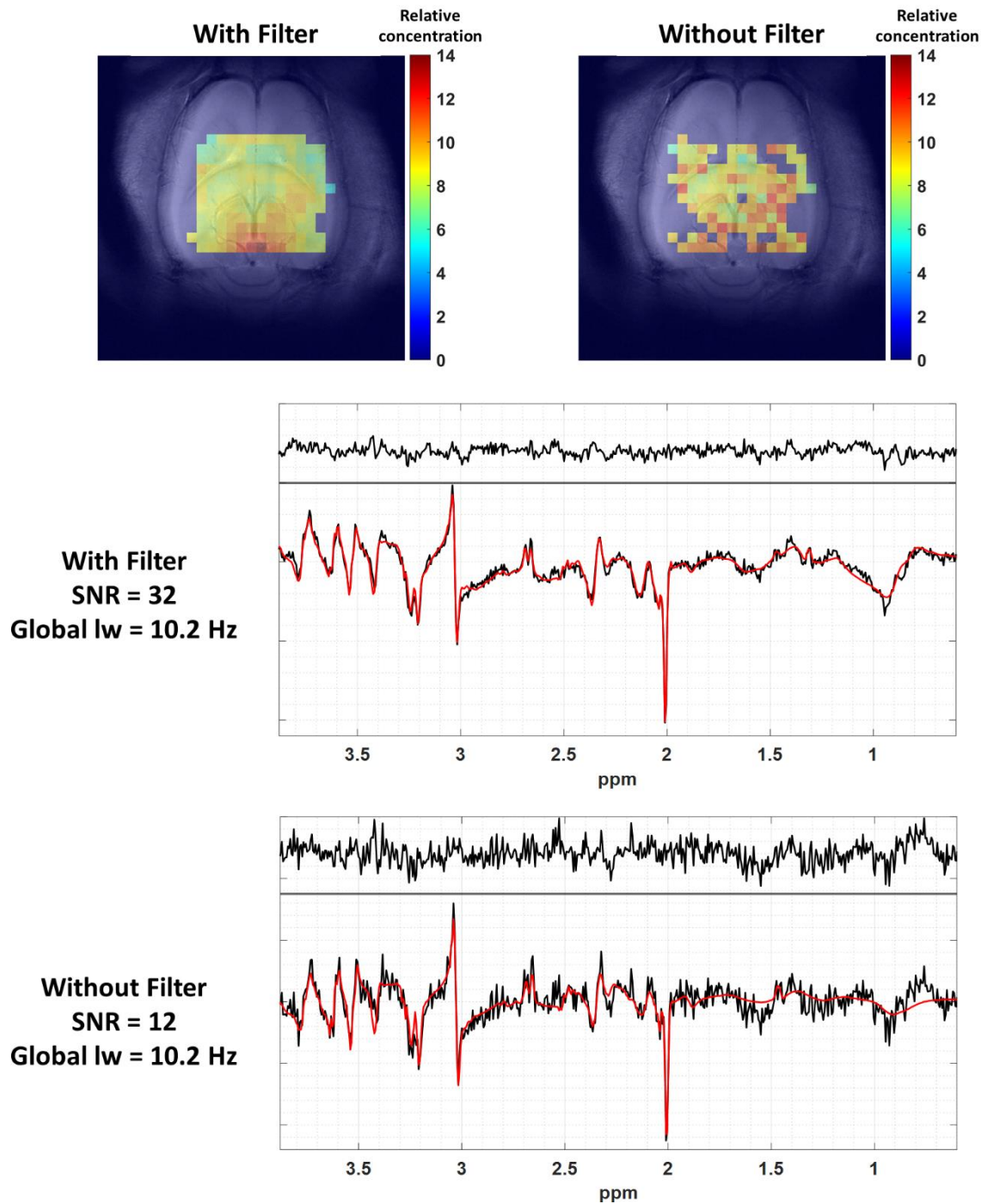

**Supplementary Figure 3**: A) Representative water $\Delta B_0$, NAA SNR, water linewidth and voxel selection after QC maps obtained after the application of the water power map on the full $^1$H-FID-MRSI matrix acquired with 1 average (QC - semi-automatic quality control applied in step 7 of the MRSI processing pipeline). Only voxels located within the brain, as determined from the water power mask, are kept, while in the QC map the remaining "bad" quality voxels are further removed during the QC steps, see part B).

B) Example of "bad" quality voxels removed by the semi-automatic quality control presented in step 7 of the MRSI processing pipeline. As can be seen these voxels show the presence of brain metabolites and not only noise, no lipid contamination is observed, however broad linewidths are visible.

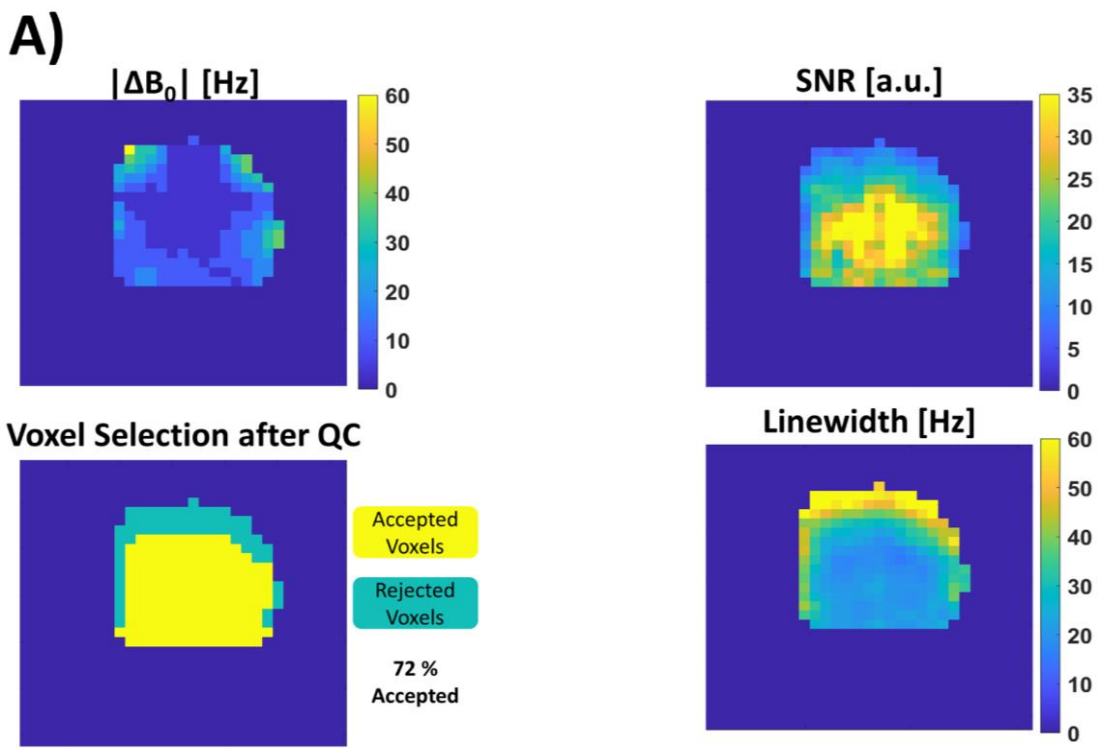

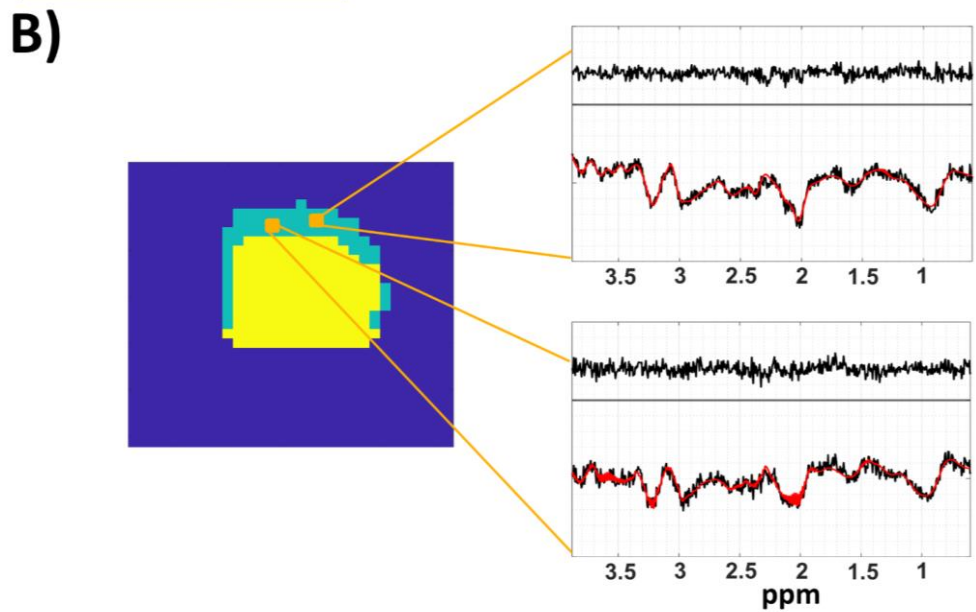

**Supplementary Figure 4:** Representative CRLB maps obtained from the LCModel quantification results of the data acquired in one rat with one average (on the left) and two averages (on the right) without QC applied. The maps were superimposed to the corresponding anatomical image using the *MRS4Brain toolbox.* The scales correspond to the CRLB limit set through the study (CRLB<30%). No interpolation was used on the metabolic maps.

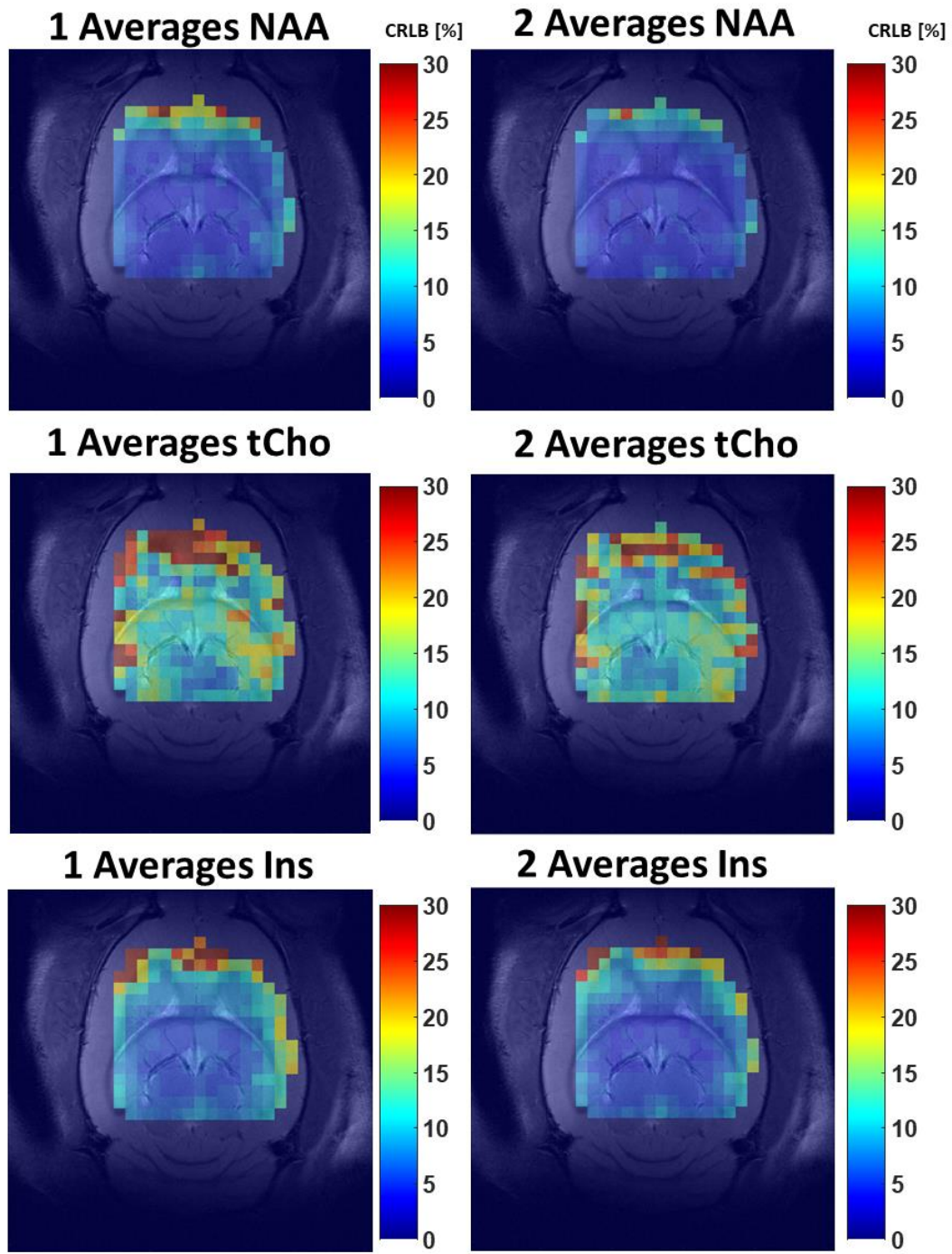

**Supplementary Figure 5**: Representative CRLB maps obtained from the LCModel quantification results of the data acquired in one rat with AD=1.3 ms (on the left) and AD= 0.94 ms (on the right), 1 average and without QC applied. The maps were superimposed to the corresponding anatomical image using the *MRS4Brain toolbox*. The scales correspond to the CRLB limit set through the study (CRLB<30%). No interpolation was used on the metabolic maps.

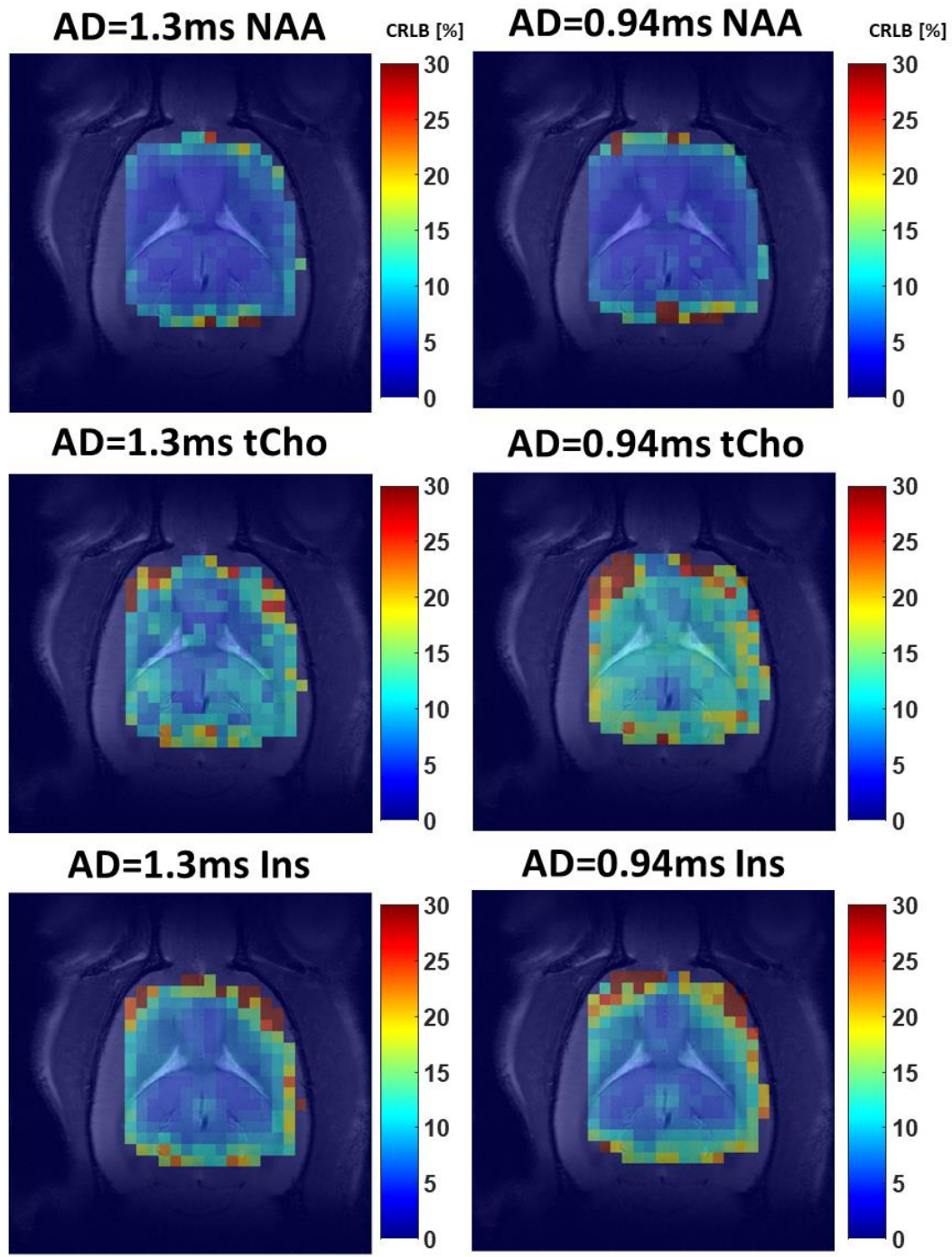

**Supplementary Figure 6**: Representative screenshot of the *MRS4Brain toolbox* incorporating the pipeline to process, quantify, display and evaluate MRSI data from a Bruker MR scanner. Version v.0.1 of the software runs in MATLAB and is available here: https://github.com/AlvBrayan/MRS4Brain-toolbox .

Additional features not highlighted in the manuscript are available in the MRSI processing pipeline: possibility to process X-nuclei MRSI data sets, change or create new Control files for LCModel fitting based on the parameters used herein for quantification (Preferences tab), open PDF files available in the acquisition folder, process multi-slice MRSI data sets, customize the specific processing techniques to be employed in the pipeline (Lipid suppression, Fillgaps, Non-Cartesian reconstruction), merge brain regions based on the automatic segmentation, calculate the brain volumetry, display concentration maps and information (mean and standard deviation). At the current moment, features with regards to reconstruction are still in the process of being added to the GUI and will be made available later on GitHub.

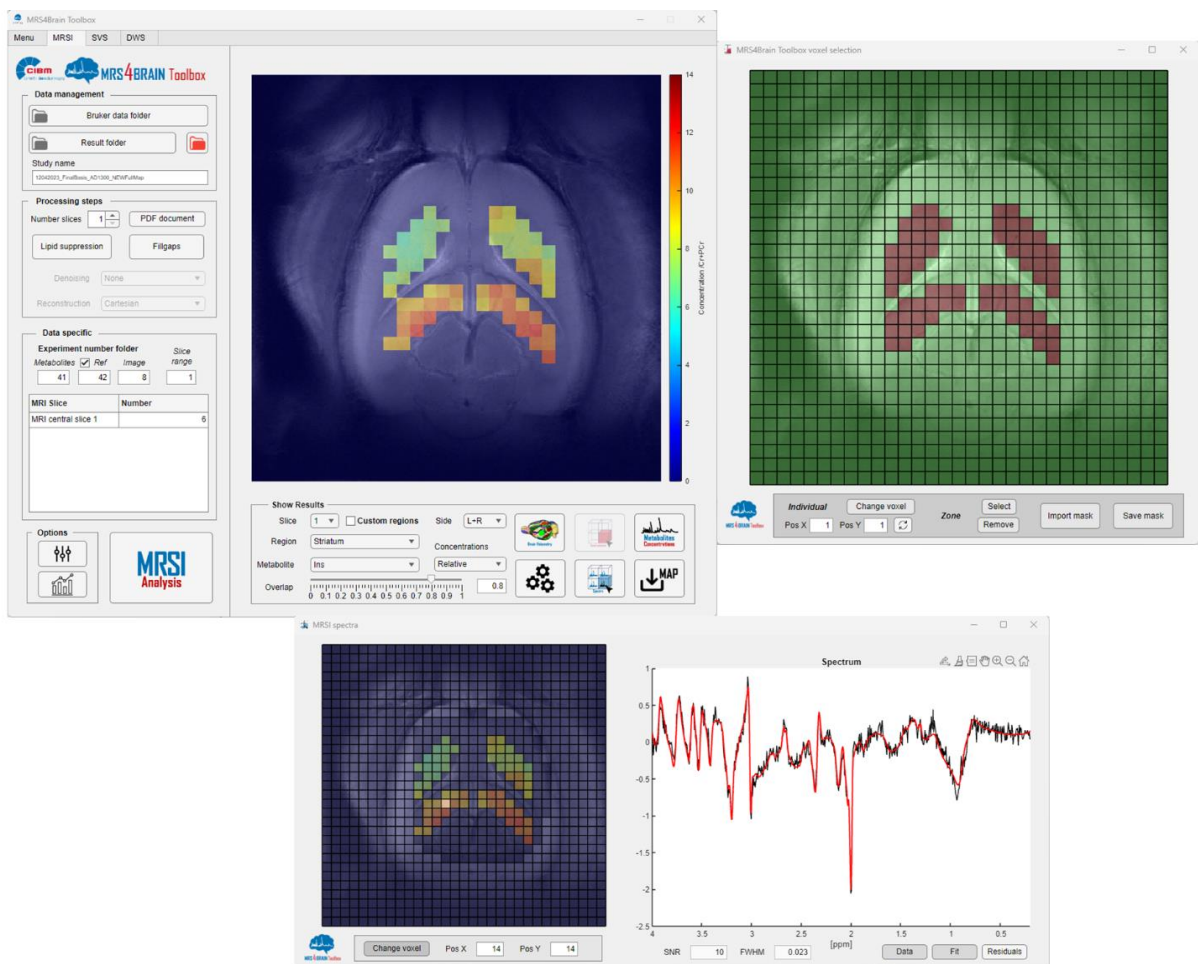

**Supplementary Figure 7:** Representative metabolic maps obtained from the LCModel quantification results of the data acquired in one rat with AD=1.3 ms (on the left) and AD= 0.94 ms (on the right), 1 average with QC applied. The brain coverage is dictated by the position of the saturation slabs, by the position and coverage of the quadrature surface coil and by the QC criteria. A tendency for decreased concentrations was observed for AD of 0.94 ms. The scales correspond to LCModel outputs when referenced to tCr by setting its concentration to 8 mmol/kg$_{ww}$. No interpolation was used on the metabolite maps. The metabolic maps were superimposed to the corresponding anatomical image using the MRS4Brain toolbox

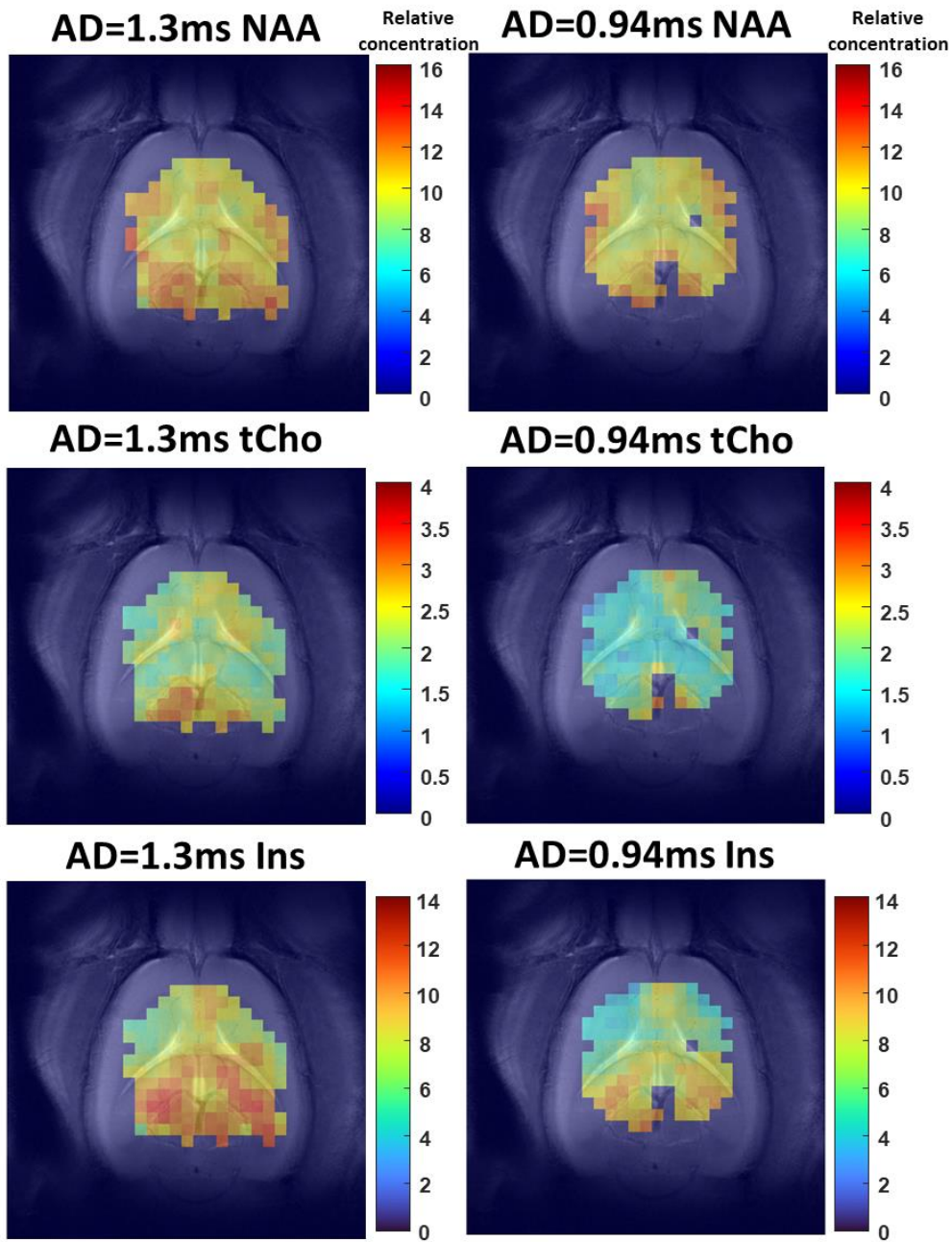

.

**Supplementary Figure 8:** Two compartment phantom $^1$H-FID-MRSI vs. PRESS-$^1$H-MRSI. **(up)** The water map provided by the jMRUI MRSI tool resulting from the PRESS-MRSI acquisition is shown on the left while the water map resulting from the $^1$H-FID-MRSI acquisition is shown on the right. The T$_2$-weighted Turbo-RARE images of the phantom in the coronal (centered on the MRSI slice) and axial positions are shown in the middle. **(down)** Spectra from three different positions (locations) in the matrix acquired with the two methods (PRESS and FID) are displayed for comparison (always the identical position for both sequences).

The "pure" Cr spectra were measured from the nominal voxels located in the Cr compartment, while the spectra from the Glu compartment (on the border with Cr compartment) contained a very minimal Cr contamination. This small Cr signal most probably comes from contamination of the Glu compartment during phantom preparation, rather than from the imperfect localization, also supported by the water maps. Of note, the Glu compartment showed a higher water residual which can further be removed in the processing step using HSVD.

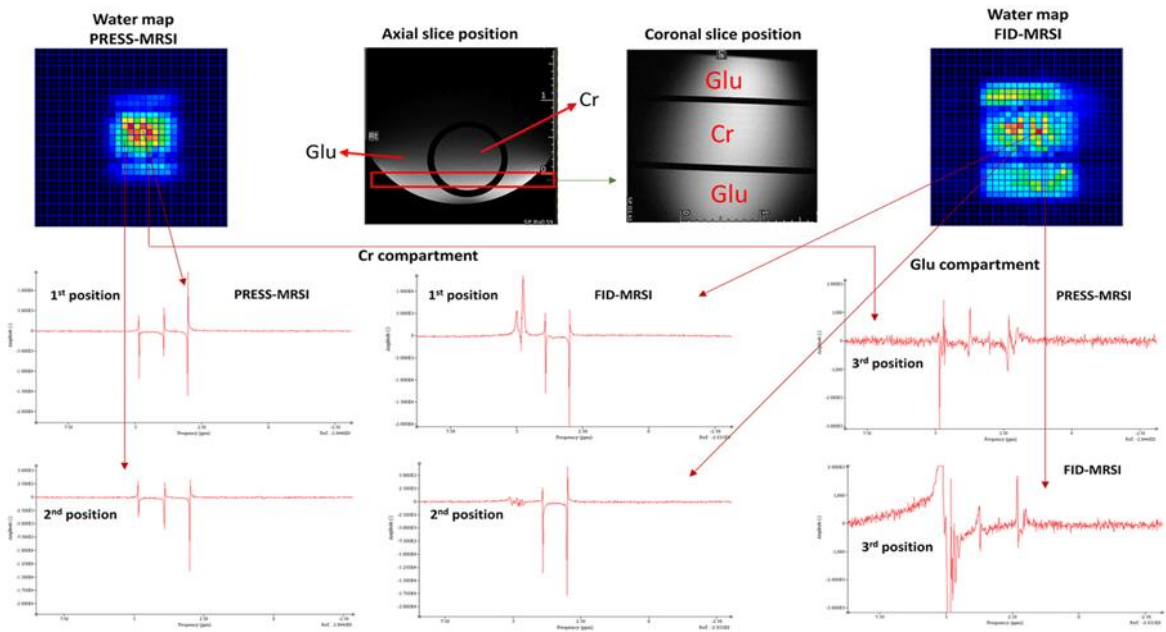

**Supplementary Figure 9:** Example of LCModel fit, using the simulated metabolite basis-set and in-vivo measured MM (Mac), showing an in-vivo spectrum at AD = 1.3 ms (blue), the LCModel fit (red), fit residual, the estimated spectra of individual metabolites and MM included in the basis-set.

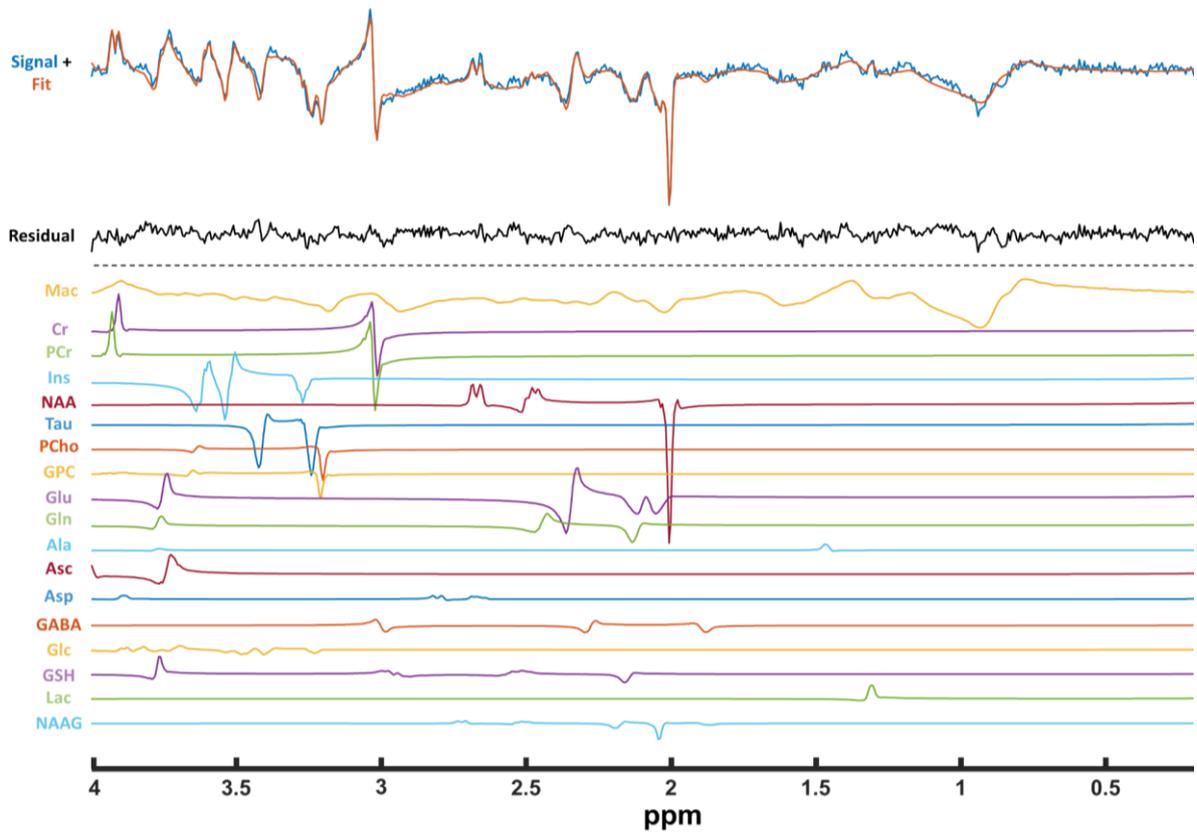

**Supplementary Figure 10:** Comparative coefficient of variation maps (%) of NAA, tCho and Ins for 1 & 2 averages (left column), 1.3 & 0.94 ms (central column) and the three time points used in the reproducibility test (0h,2h,4h; right column). The coefficient of variation was measured as the ratio per voxel of one standard deviation over the mean of the relative concentration estimated after quantification.

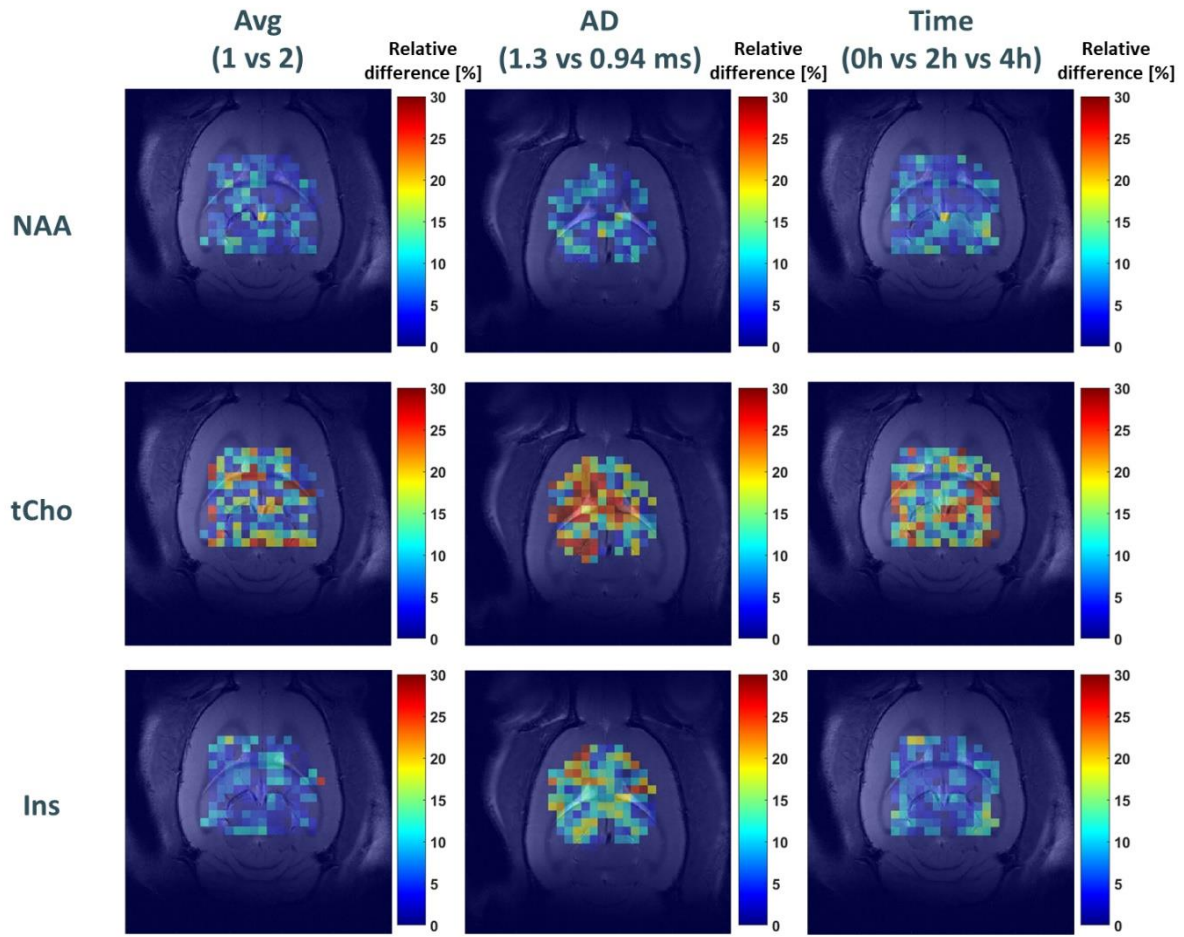

**Supplementary Figure 11:** 2D and 1D profiles plot of the Point Spread Function (PSF), without and with the in-plane Hamming filter described in section 2.1. applied. On the 1D profiles, boundaries of the signal emitting voxel (voxel 0) are represented in red dashed lines. The width at half maximum is calculated in terms of voxel size.

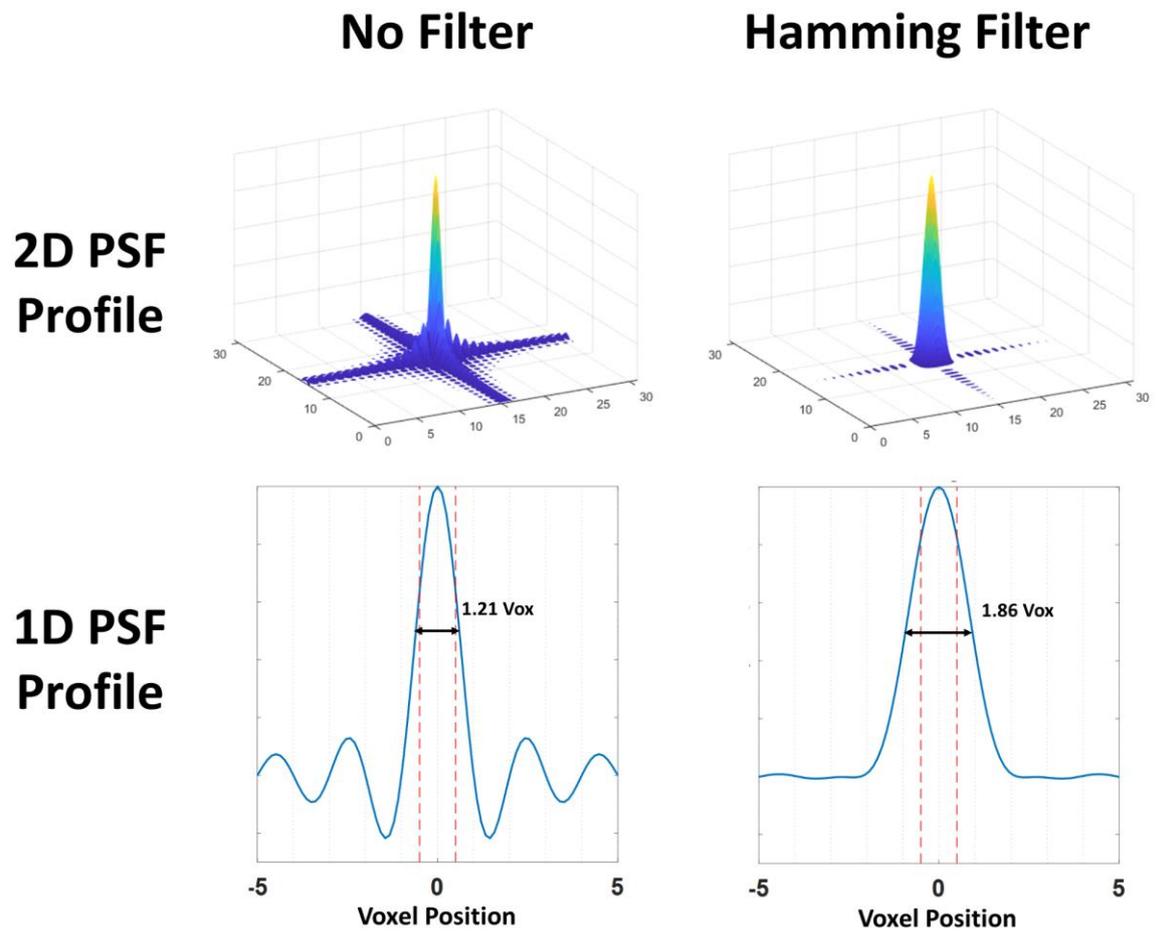

**Supplementary Table 1:** Minimum reporting standards in MRS

## Hardware

| | |
|---|---|
| **Field strength** | 14.1T |
| **Manufacturer** | Bruker |
| **model (software)** | (Paravision 360 V1.1. & V3.3.) |
| **rf coil** | $^1$H-quadrature surface head coil |
| **additional hardware** | N/A |

## Acquisition

| | |
|---|---|
| **pulse sequence** | FID-MRSI |
| **volume of interest (voi)** | Rodent: Brain |
| **nominal voi size** | 0.77 × 0.77 × 2 mm$^3$ |
| **repetition time TR & Acquisition delay AD** | TR = 813ms / AD = {1.3, 0.94} ms |
| **number of excitations per spectrum** | 1 & 2 averages |
| **additional parameters** | 24 × 24 × 2 mm$^3$ FOV; matrix size 31×31; no acceleration factor; Cartesian *k*-space sampling<br><br>In-plane Hamming *k*-space filter in each dimension post processing<br><br>Six saturation slabs |
| **water suppression method** | VAPOR |
| **shimming method** | Bruker MAPSHIM, first in an ellipsoid covering the full brain and further in a volume of interest centered on the MRSI slice.; <30 Hz for $H_2O$ resonance |
| **triggering or motion correction** | N/A |

## DATA ANALYSIS

| | |
|---|---|
| **Analysis Software** | LCmodel (Version 6.3-1N) |
| **Processing step deviating from reference** | Custom Basis-Set (for both AD=1.3ms and AD=0.94ms) |
| | Control files provided with the *MRS4Brain toolbox* |
| **output measure** | Ratios to total Creatine |
| **quantification reference** | Basis-set including: alanine, aspartate, ascorbate, creatine, phosphocreatine, γ-aminobutyrate, glutamine, glutamate, glycerophosphocholine, glutathione, glucose, inositol, N-acetylaspartate, N-acetylaspartylglutamate, phosphocholine, phosphoethanolamine, lactate, taurine simulated using NMR ScopeB. Macromolecules acquired in-vivo with double inversion recovery FID-MRSI. |

## Data quality

| | |
|---|---|
| **reported variables** | SNR (reference to NAA) and linewidths (reference to water) both reported <br><br> Global linewidths estimated by LCModel |
| **Data exclusion criteria** | LCModel SNR > 4 and <br><br> LCModel FWHM < 1.25*LCModel |
| **quality measures of post-processing model fitting** | CRLB < 30% |
| **sample spectrum** | Figure 2/ Supplementary Figures 2,3,6,7,8 |

**Supplementary Table 2:** CRLB quantitative assessment of the relative concentration estimates (tCr as reference), over the whole slice using 1 and 2 averages (mean of the mean, over 7 measurements in 6 rats, per average) at AD=1.3 ms. Mean values and standard deviation (calculated with error propagation from tCr CRLB) were computed after application of the semi-automatic quality control (after step 7).

| 1 Averages | | 2 Averages | |
|---|---|---|---|
| CRLB [%] | | CRLB [%] | |
| NAA | 7.93 ± 0.52 | NAA | 6.98 ± 0.53 |
| Gln | 11.84 ± 0.69 | Gln | 10.01 ± 0.82 |
| Glu | 8.39 ± 0.53 | Glu | 7.33 ± 0.54 |
| Ins | 9.51 ± 0.73 | Ins | 8.36 ± 0.76 |
| Tau | 11.22 ± 2.32 | Tau | 9.96 ± 2.62 |
| GPC+PCho | 12.17 ± 1.21 | GPC+PCho | 11.38 ± 1.28 |
| NAA+NAAG | 7.85 ± 0.47 | NAA+NAAG | 6.94 ± 0.48 |

**Supplementary Table 3:** CRLB quantitative assessment of the relative concentration estimates (tCr as reference), over the whole slice using AD = 1.3 ms and 0.94 ms (mean of the mean, over 4 rats, per AD) with 1 average. Mean values and standard deviation (calculated with error propagation from tCr CRLB) were computed after application of the semi-automatic quality control (after step 7).

| AD=1.3ms | | AD=0.94ms | |
|---|---|---|---|
| CRLB [%] | | CRLB [%] | |
| NAA | 8.12 ± 0.68 | NAA | 7.21 ± 0.57 |
| Gln | 12.04 ± 0.88 | Gln | 11.63 ± 1.13 |
| Glu | 8.85 ± 0.67 | Glu | 8.02 ± 0.58 |
| Ins | 9.66 ± 1.01 | Ins | 8.97 ± 0.99 |
| Tau | 11.20 ± 1.98 | Tau | 10.27 ± 1.67 |
| GPC+PCho | 10.91 ± 1.15 | GPC+PCho | 13.53 ± 1.38 |
| NAA+NAAG | 8.08 ± 0.62 | NAA+NAAG | 7.24 ± 0.49 |

**Supplementary Table 4:** Quantitative reproducibility assessment in two brain regions using AD = 1.3 ms and 0.94 ms (mean of the mean over 4 rats, per AD) with 1 average. Brains regional differences are marked with *,**. Statistically significant differences between ADs for Ins (13%** hippocampus and

16%* cortex+striatum), tCho (16%** hippocampus and 16%* cortex+striatum) and Tau (9%*, only for hippocampus), not highlighted in the table.

| */tCr [mmol/kg$_{ww}$] | AD=1.3ms | | */tCr [mmol/kg$_{ww}$] | AD=0.94ms | |
| --- | --- | --- | --- | --- | --- |
| | Hippocampus | Striatum + Cortex | | Hippocampus | Striatum + Cortex |
| NAA | 10.18 ± 0.40 | 9.83 ± 0.50 | NAA | 9.62 ± 0.54 | 9.92 ± 0.50 |
| Gln (*) | 5.38 ± 0.37 | 6.43 ± 0.43 | Gln (*) | 5.21 ± 0.42 | 5.99 ± 0.44 |
| Glu | 11.22 ± 0.42 | 11.41 ± 0.55 | Glu | 10.77 ± 0.51 | 11.36 ± 0.55 |
| Ins (**) | 10.11 ± 0.47 | 7.98 ± 0.60 | Ins (**) | 8.78 ± 0.51 | 6.73 ± 0.71 |
| Tau (*) | 7.29 ± 0.34 | 8.20 ± 0.61 | Tau | 6.65 ± 0.37 | 7.32 ± 0.60 |
| GPC+PCho (*) | 1.87 ± 0.10 | 2.18 ± 0.17 | GPC+PCho (*) | 1.58 ± 0.12 | 1.84 ± 0.17 |
| NAA+NAAG | 11.57 ± 0.51 | 11.03 ± 0.63 | NAA+NAAG | 10.90 ± 0.58 | 11.15 ± 0.79 |